\documentclass[10pt,journal,comsoc]{IEEEtran}
\usepackage{graphicx}
\usepackage{epstopdf}
\usepackage[T1]{fontenc}
\usepackage[latin9]{inputenc}
\usepackage{float}
\usepackage{calc}
\usepackage{mathrsfs}
\usepackage{amsmath}
\usepackage{amsthm}
\usepackage{xcolor}
\usepackage{ulem}

\usepackage[unicode=true,
 bookmarks=true,bookmarksnumbered=true,bookmarksopen=true,bookmarksopenlevel=1,
 breaklinks=false,pdfborder={0 0 0},pdfborderstyle={},backref=false,colorlinks=false]
 {hyperref}
\hypersetup{pdftitle={Your Title},
 pdfauthor={Your Name},
 pdfpagelayout=OneColumn, pdfnewwindow=true, pdfstartview=XYZ, plainpages=false}

\makeatletter
\providecommand{\tabularnewline}{\\}
\floatstyle{ruled}
\newfloat{algorithm}{tbp}{loa}
\providecommand{\algorithmname}{Algorithm}
\floatname{algorithm}{\protect\algorithmname}
\let\oldforeign@language\foreign@language
\DeclareRobustCommand{\foreign@language}[1]{%
  \lowercase{\oldforeign@language{#1}}}
\theoremstyle{plain}
\newtheorem{lem}{\protect\lemmaname}
\theoremstyle{plain}
\newtheorem{thm}{\protect\theoremname}
\usepackage[caption=false,font=footnotesize]{subfig}

\@ifundefined{showcaptionsetup}{}{%
 \PassOptionsToPackage{caption=false}{subfig}}
\usepackage{subfig}
\makeatother

\providecommand{\lemmaname}{Lemma}
\providecommand{\theoremname}{Theorem}

\ifCLASSOPTIONcompsoc
  \usepackage[nocompress]{cite}
\else
  \usepackage{cite}
\fi

\ifCLASSINFOpdf

\else

\fi

\begin{document}
\normalem 

\title{Connectivity-Aware Contract for Incentivizing IoT Devices in Complex
Wireless Blockchain}

\author{
  Weiyi~Wang,~Jin~Chen,~Yutao~Jiao,~Jiawen~Kang,~Wenting Dai,~and~Yuhua Xu

\thanks{Weiyi Wang, Jin Chen, Yutao Jiao, and Yuhua Xu are with the College of Communications Engineering, 
Army Engineering University of PLA, Nanjing 210007, China.}
\thanks{Jiawen Kang is with the School of Automation, Guangdong University of Technology, Guangzhou 
510006, China, and also with the Key Laboratory of Intelligent Information Processing and System 
Integration of IoT, Ministry of Education, Guangzhou 510006, China.}
\thanks{Wenting Dai is with the School of Communications and Information Engineering, Nanjing 
University of Posts and Telecommunications, Nanjing 210003, China.}
}

\IEEEtitleabstractindextext{%
\begin{abstract}
  Blockchain is considered the critical backbone technology for 
secure and trusted Internet of Things (IoT) in the future 6G network. 
However, deploying a blockchain system in a complex wireless IoT network 
is challenging due to the limited resources, complex wireless environment,
and the property of self-interested IoT devices.
The existing incentive mechanism of blockchain is not compatible with the wireless IoT network. 
In this paper, to incentivize IoT devices to join the construction 
of the wireless blockchain network, we propose a multi-dimensional contract to 
optimize the blockchain utility while addressing the issues of adverse selection and 
moral hazard. Specifically, the proposed contract considers the IoT device's 
hash power and communication cost and especially explores the connectivity of devices 
from the perspective of complex network theory. We investigate the 
energy consumption and the block confirmation probability 
of the wireless blockchain network via simulations under varied network sizes and average link probability. 
Numerical results demonstrate that our proposed contract mechanism is feasible, achieves 35\% more 
utility than existing approaches, and increases utility by 4 times compared with the original PoW-based 
incentive mechanism.
\end{abstract}

}

\maketitle

\IEEEdisplaynontitleabstractindextext
\IEEEpeerreviewmaketitle

\section{Introduction}

\IEEEPARstart{B}{lockchain}, also known as distributed ledger technology 
(DLT), has attracted substantial interest. Due to the distributed, cryptographic,
immutable, token, and decentralized characteristics, blockchain shows 
excellent potential to be a critical technology for securing future 6G networks.
Nowadays, blockchain is not only applied in the field of finance 
(e.g., cryptocurrency \cite{Nakamoto2017}) but also in the Internet of Things (IoT) \cite{IoTJ}.
The authors in \cite{Lwin2020} used blockchain to record the 
node's trust value, which is utilized to differentiate the malicious node. 
Similarly, the authors in \cite{Kang2018} proposed 
a consensus management scheme to ensure secure miner selection in the Internet 
of Vehicles (IoV). Blockchain was used to store reputation and provide 
trust in the IoV network. Most current works focus on the 
blockchain-based scenario and assume that the blockchain has been deployed and
well-operated. However, the IoT network usually has a complex wireless environment, which influences the blockchain 
performance and challenges deploying the wireless blockchain. The authors in 
\cite{wirelessPBFT} analyzed the impact of transmission power on blockchain consensus in 
the wireless network. In \cite{Xiao2020}, the authors derived that the device's connectivity 
determines the forking rate.
 
Nevertheless, only some works consider wireless 
communication when using blockchain in IoT networks. On the one hand, the IoT device's 
complex wireless network topology significantly affects blockchain performance.
On the other hand, the limited energy of IoT devices reduces their willingness to join the blockchain.
As the most widely used consensus mechanism, Proof-of-Work (PoW) consensus requires participants to 
consume energy and resources to solve the hash puzzle. Unfortunately, most IoT devices have limited 
battery capacity, restricted communication and computing capabilities, and lack the motivation to 
participate in the blockchain network. Besides, existing PoW-based incentive mechanisms allocate 
the same rewards for devices while neglecting their heterogeneity.
During block propagation, the IoT devices' transmission power influence the communication 
efficiency, and the network connectivity affects the propagation 
delay. Thus, there needs to design an adequate incentive mechanism 
for deploying blockchain in the IoT network.

To encourage IoT devices to join the wireless blockchain network, we need to 
consider the above heterogeneous and limited capabilities of IoT devices and design a desirable 
incentive mechanism in reward allocation. The incentive mechanism also needs to assess the quality 
of the work. For one thing, 
the IoT device may work passively, and for another, the block is at risk of confirmation 
failure due to the forking or communication outage. Therefore, the incentive mechanism 
should have the following properties: 1) Modeling the impact of both computing 
and wireless communication factors on blockchain performance and IoT devices' energy 
consumption; 2) Reflecting the IoT devices' actual capabilities and preferences; 
3) Distributing rewards based on the quality of task completion.

Contract theory is an effective method to address the issues of information asymmetry 
and passive behavior. Since wireless blockchain performance is related to multiple 
factors, we design a multi-dimensional contract to incentivize heterogeneous IoT devices 
to maximize blockchain utility. In our proposed contract, the blockchain utility is designed 
based on the duration of participating in the blockchain network, and the IoT device's task is to generate 
blocks. The more IoT devices join the blockchain, and the more blocks are proposed, the more utility 
the blockchain obtains. We respectively analyze the impact of hash power $c$, transmission power $p$, and 
connectivity $c$ on block propagation and energy consumption. 
The main contributions of this paper can be summarized as follows:
\begin{itemize}
\item We design a novel multi-dimensional contract model addressing both \emph{adverse selection} 
and \emph{moral hazard} for maximizing the blockchain utility in the wireless IoT network. The 
\emph{adverse selection} is used to reveal the actual capabilities of wireless IoT devices and their 
preference. The \emph{moral hazard} measures the quality of the task completion.
\item To incentivize IoT devices to join the wireless blockchain, 
our contract jointly considers IoT devices' hash power, transmission power, and connectivity. 
We analyze how these factors determine energy consumption and block confirmation probability.
\item Particularly, we characterize and verify the logarithmic relationship between the connectivity 
and confirmation probability by the experimental results. We also find that the device's connectivity 
significantly impacts the block confirmation probability under varied network sizes and average 
link probability. 
\item The numerical results demonstrate that the proposed contract efficiently
incentivizes the IoT devices, improves blockchain utility by {35\%} compared with the contract 
with adverse selection and increases utility by 4 times 
compared with the original PoW-based incentive mechanism.
\end{itemize}

The rest of this paper is organized as follows. Section \mbox{II}
reviews the related work. Section \mbox{III} presents the
system model and the performance metrics of the wireless blockchain network
and IoT devices. The optimal multi-dimensional contract is  proposed in 
Section \mbox{IV}. Section \mbox{V} presents the numerical results to validate 
the contract's feasibility and effectiveness. Finally, Section 
\mbox{VI} concludes the paper. Table 1 lists the main notations of this paper.
\begin{table}[htbp]
\caption{Main Notations}

\centering{}%
\begin{tabular}{ll}
\hline 
Notation & Definition\tabularnewline
\hline 
$Z$ & Number of IoT devices\tabularnewline
$L$ & Number of IoT devices' types\tabularnewline
$P_{l}$ & Average link probability of the network\tabularnewline
$Q$ & Probability distribution function for the type\tabularnewline
$G_{i}$ & Confirmation probability of the type $i$ device\tabularnewline
$F_{i}$ & Energy cost of the type $i$ device\tabularnewline
$P_{h}$ & Block confirmation probability determined by hash power\tabularnewline
$P_{c}$ & Block confirmation probability determined by connectivity\tabularnewline
$P_{out}$ & Communication outage probability\tabularnewline
$a$ & Block size\tabularnewline
$h_{i}$ & Hash power of the type $i$ device\tabularnewline
$c_{i}$ & Connectivity of the type $i$ device\tabularnewline
$p_{i}$ & Transmission power of the type $i$ device\tabularnewline
$\lambda$ & Converted type of IoT device\tabularnewline
$e_{i}$ & Number of blocks proposed by the type $i$ device\tabularnewline
$s_{i}$ & Salary for the type $i$ device\tabularnewline
$B_{i}$ & Bonus for the type $i$ device\tabularnewline
$b_{i}$ & Unit bonus for the type $i$ device\tabularnewline
$\omega_{i}$ & Contract item for the type $i$ device\tabularnewline
$\theta$ & Time cost coefficient\tabularnewline
$\varepsilon$ & Yield coefficient\tabularnewline
$\gamma$ & Energy cost coefficient\tabularnewline
$r$ & Transmission rate\tabularnewline
$\tau$ & Average block interval\tabularnewline
\hline 
\end{tabular}
\end{table}

\section{Related Work}
\subsection{Blockchain for IoT}

Since Satoshi Nakamoto proposed bitcoin in 2008 \cite{Nakamoto2017},
blockchain has attracted lots of attention. Incipiently, blockchain
is used in the cryptocurrency field, and the birth of Ethereum extended
the application of blockchain. Ethereum introduces the
smart contract \cite{Eth}, where the code will run automatically while satisfying
the input criteria. Currently, there are many works combining the
IoT and blockchain. The authors in \cite{blot} analyzed the opportunities
and challenges of applying blockchain in IoT. In \cite{Feng},
the authors proposed a distributed consensus protocol for the Internet of Vehicles
(IoV) and enhanced the system's stability. However, adding a negative vote to the 
consensus mechanism may reduce security. The authors in \cite{Lwin2020}
proposed the blockchain-based trust management system in mobile ad-hoc
networks, where the computing complexity of the consensus algorithm
is decreased. But the algorithm can't deal with the situation when the neighbors of the 
node are all malicious. A reputation-based routing method using blockchain is
provided in a mobile ad-hoc network \cite{9027450}. In this protocol, nodes select 
the routing path to forward, considering the length and reputation.  In \cite{Kang2018},
the authors investigated the blockchain-enabled IoV to enhance
network security. They calculate reputations for every participant and vote miners based on reputations. 
The authors in \cite{Sun2019} designed a blockchain-based
wireless IoT model and provided throughput analyses. They derived
the optimal full node deployment and analyzed the model's security
under attacks. In \cite{Cao2019}, the authors studied the existing
Direct Acyclic Graph (DAG) consensus protocols designed for IoT. The
research shows that the Tangle \cite{Tangle} and Hashgraph \cite{Hashgraph}
are more suitable for IoT than Proof of Work (PoW) and Proof of Stake
(PoS). However, the security of DAG-based consensus needs to be verified.

Recently, some research has been focusing on the protocol 
analysis and resource allocation of the blockchain network. The authors 
investigated the information propagation in
bitcoin and presented three schemes to improve the propagation
delay \cite{Decker2013}. Though the third scheme decreases the propagation delay, 
it has a high demand on the communication bandwidth. In \cite{bpv}, the authors studied the 
impact of mobility on block propagation in the vehicular network. They gave the 
closed-form expression of the single-block propagation time and found that high mobility and 
connectivity speed up the block propagation. The authors in \cite{Shahsavari2020} proposed 
a theoretical model for analyzing block propagation in the bitcoin network. They modeled the performance
using a random graph model and derived the explicit equations of block
propagation delay. The energy consumption of blockchain was discussed
and explored in \cite{Ghosh2020} and \cite{Sedlmeir2020}. But they mainly focus 
on the energy consumption of computing, neglecting communication. The authors in 
\cite{Huang2021} proposed a resource allocation scheme to minimize the cost of access and storage in 
a blockchain-based edge computing network. The proposed method applies to the 
scenario where the network topology changes slowly. There are also some works aware of the impact of
communication on blockchain performance. The authors in \cite{Wei}
investigated how communication reliability affects the PoW consensus
mechanism. They analyzed blockchain security from a new perspective
but just gave a simple qualitative analysis. While guaranteeing
the safety of the Practical Byzantine Fault Tolerance (PBFT) protocol,
the authors in \cite{PBFT} showed the minimum transmission power of
nodes in the wireless network. But the paper just considers the transmission power 
and lacks analyses of other wireless factors. In \cite{Xiao2020},
the authors proposed a probability analytic model that evaluated blockchain
security considering the node's connectivity and computing power.
The authors in \cite{Lei} investigated the communication
resource consumption in the wireless blockchain network and analyzed the communication
resources required by different consensus protocols. This paper makes a qualitative 
analysis of various consensus agreements but lacks in-depth exploration. 
The authors combined the blockchain and the access protocol to
investigate the blockchain throughput in \cite{Li2021}. They assumed that there were only block messages 
in channels, which is unreasonable.  
The authors  discussed blockchain-enabled wireless applications and 
proposed a wireless blockchain middleware architecture \cite{li2021blockchain}. They provided 
several research directions for wireless blockchain. In \cite{gill2022ai}, authors analyzed the 
function of blockchain for next-generation computing. Their work showed that blockchain has great 
potential in cloud/fog/edge computing. The authors in \cite{hafid2022tractable} investigated the 
Sybil attacks in sharding-based blockchain protocols and gave a tractable probabilistic approach to 
evaluate blockchain security.
However, fewer works consider the impact of 
communication factors on blockchain performance, especially in a complex wireless environment.

\subsection{Incentive Mechanism for Blockchain Network}

Contract theory is a typical mechanism design method in real-world
economics and has been widely used to model the relationship between
employers and employees \cite{Contract}. 
Applying the mechanism design approach 
to the wireless IoT network has been investigated widely \cite{Gao2011}, \cite{Jiao3}, and \cite{Lim2020}.
Moreover, integrating blockchain and incentive mechanisms is also
a hot topic. 
In Bitcoin and Ethereum, miners consume their computing power to solve the hash puzzle and obtain the 
corresponding tokens \cite{Nakamoto2017} and \cite{Eth}. Besides, miners could select transactions to place 
on the block to earn transaction fees, which is considered a first-price auction. In \cite{roughgarden2020transaction}, 
the author analyzed the EIP-1559 of Ethereum. This mechanism divides the transaction fee into base fee and tips, 
where the base fee is paid to miners, and the tips are burned. In Storj \cite{wilkinson2014storj}, participants contribute 
their storage and bandwidth to obtain tokens. Not all incentive mechanisms provide rewards. In 
Casper \cite{buterin2020incentives}, devices will be punished if they sign on conflicting blocks.
In \cite{Jiao2019}, the authors
utilized the auction to allocate computing resources in blockchain
networks. They proposed two schemes to satisfy miners' demands flexibly. The authors in \cite{Shi2021} summarized 
blockchain-based auction applications and expounded the auction-based solutions for
blockchain enhancement. The authors in \cite{Lijing} studied how
to determine the deposit threshold using contract theory in the sharded
blockchain. They designed a one-dimensional contract to decide different deposit thresholds for 
heterogeneous users, which provides more opportunities for participants with low stake values.

However, the above works do not tackle the problem of incentive mechanism 
design for blockchain deployment in the complex wireless network. 
In the IoT network, the communication and computing capabilities of devices both decide the block 
confirmation. Nevertheless, the traditional incentive mechanism just 
allocates token rewards according to the computing capacity, which is inappropriate for the wireless 
blockchain network. In addition, the classical PoW-based incentive mechanism provides the same rewards 
for devices. Under this scheme of allocation policy, powerful devices may dissatisfy with the received 
rewards since they consume more resources and low-energy devices have no motivation to join the blockchain 
due to uncertain incentives. 
Therefore, we design a multi-dimensional contract containing IoT devices' cost and confirmation 
probability to incentivize them in the complex wireless blockchain network. Our contract provides 
a guaranteed reward for participants.

\begin{figure*}[t]
\begin{centering}
\includegraphics[scale=0.45]{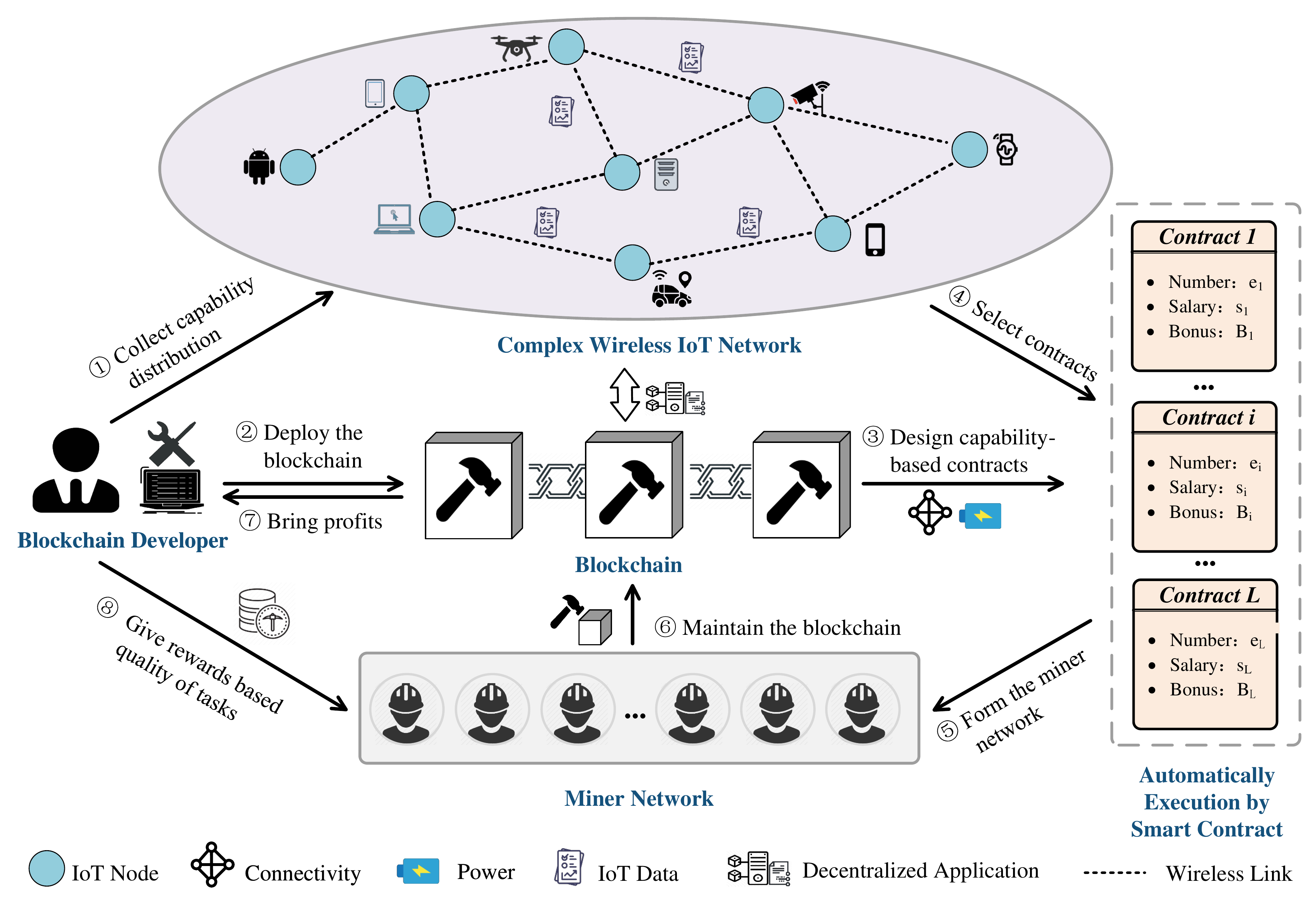}
\par\end{centering}
\caption{The integrated system of blockchain and IoT in the complex wireless network.}
\end{figure*}

\section{System Model: Incentive Mechanism for Wireless Blockchain Network in IoT}

In this section, we describe the integrated system of blockchain and
IoT in the complex wireless network. As is shown in Fig. 1, there are many 
kinds of IoT devices in the wireless network, such as mobile phones, 
laptops, UAVs, intelligent robots, etc. The blockchain developer aims to 
attract more IoT devices to participate in the wireless blockchain network. 
Nevertheless, IoT devices lack the motivation to maintain blockchain due to 
time and energy consumption. Thus, the developer uses contract 
theory to incentivize IoT devices to join the blockchain network while 
maximizing the blockchain utility. First, the blockchain developer collects the 
distribution of IoT devices' capabilities in advance to deploy the blockchain better 
and design contracts based on the information. The contracts are stored in the
blockchain in the form of smart contracts. After selecting a contract, the IoT 
devices form the miner networks and operate the wireless blockchain protocol according 
to the signed contract for securing the blockchain. Finally, the blockchain 
developer inspects the quality of tasks and sends rewards to IoT devices.

\subsection{Complex Wireless Blockchain Network}

IoT devices are heterogeneous in computing and communication capabilities 
in the complex wireless blockchain system. Additionally, IoT devices communicate with others 
via unreliable wireless links, where communication outages are common. 
Another important metric in the complex wireless network is connectivity. 
The device's connectivity reflects its communication advantage and has a significant 
impact on the block propagation delay. Deploying the blockchain in the complex wireless 
network not only takes into account the heterogeneity of the IoT devices but also the impact of 
communication factors on blockchain performance.

\begin{figure}[htbp]
  \begin{centering}
  \includegraphics[scale=0.65]{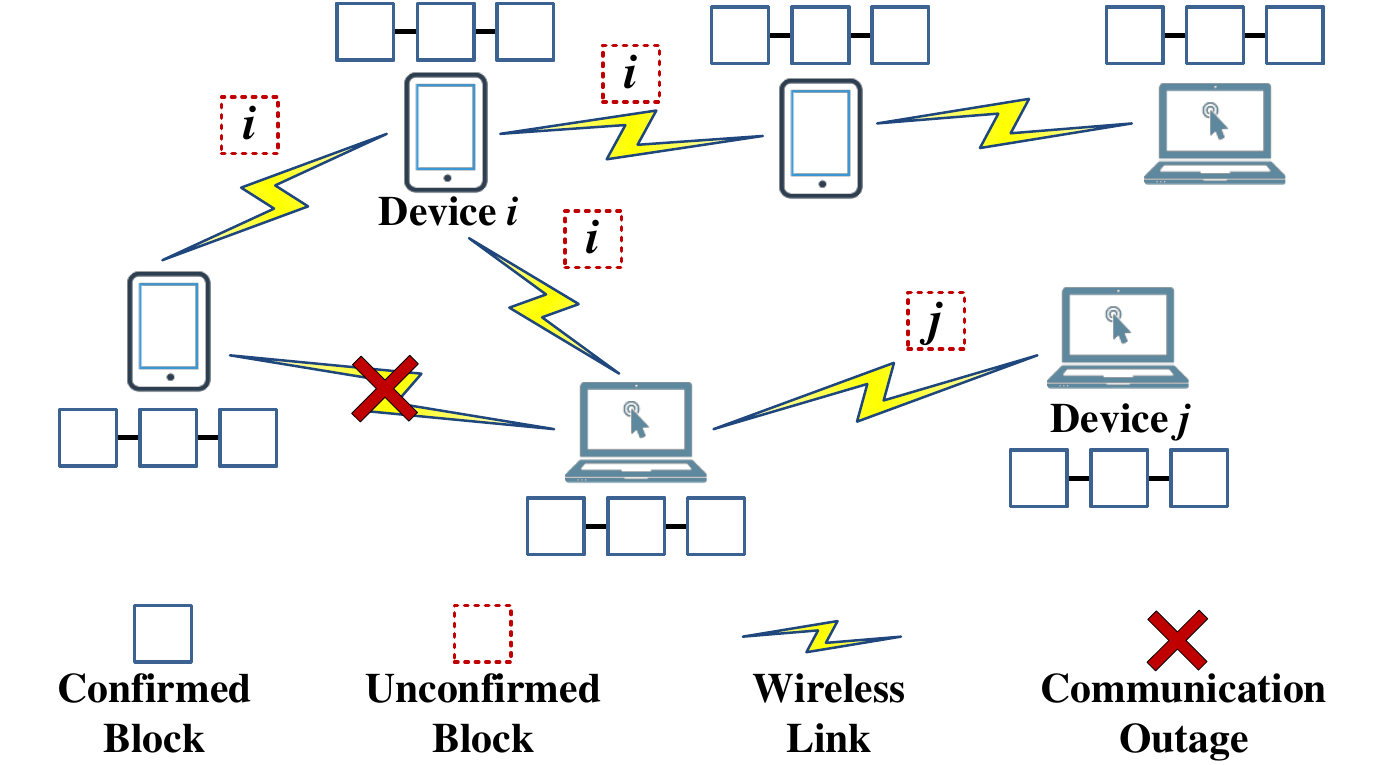}
  \par\end{centering}
  \caption{The forking event in the wireless blockchain network.}
  \end{figure}

Unlike traditional PoW-based blockchains, the hash power $h$ is the dominant factor affecting consensus. 
By contrast, communication factors have trivial effects on consensus. Nevertheless, both transmission 
power $p$ and connectivity $c$ significantly influence the block confirmation probability in the wireless 
blockchain and further influence consensus. As shown in Fig. 2, both device $i$ and device $j$ propose 
the block at height $16$. According to the longest legal chain principle, only one block will be confirmed, 
and the other will be abandoned. Device $i$ has a higher transmission power $p$ and connectivity $c$ so 
that its block is accepted faster by other devices. Thus, the block proposed by device $i$ is more likely 
to be confirmed and wins the competition of the forking. This paper considers the two-prong forking, the 
most pervasive and possible situation.

Due to the influence of the complex network, the classical PoW-based incentive mechanism is not compatible 
well. For one thing, the reward is the same for all IoT devices, regardless of cost. For another, 
the complex wireless network increases the risk of forking, which also reduces IoT devices' motivation 
to join the blockchain. Thus, we design a multi-dimensional contract $(e,R)$ to address the issues. 
The variable $e$ represents the number of blocks in the longest legal chain, and $R$ is the reward 
distributed to the IoT device. The reward $R$ is defined as follows:

\begin{equation}
R=s+B,
\end{equation}

\begin{equation}
B=be,
\end{equation}
where $s$ is the fixed salary, $B$ is the bonus, and $b$ is the unit 
bonus. The salary represents the basic income, and the bonus is related to the 
proposed blocks. The more blocks proposed, the more bonuses the IoT device obtains.

With the basic requirement of an effective contract, our proposed contract should also 
satisfy the following properties:
\begin{itemize}
\item \emph{Individual Rationality} (IR). IR condition means that
the IoT devices can obtain positive utility while signing the contract, 
which is the foundation for incentivizing the device to join the wireless blockchain.
\item \emph{Incentive Compatibility} (IC). IC condition means that every
IoT device can only achieve the maximum utility by choosing the contract
according to its actual computing and communication capability. It helps the contract designer
be aware of the true preference of IoT devices.
\end{itemize}

\subsection{IoT Device Model in Wireless Blockchain Network }

In the energy-constrained wireless network, energy is precious
for IoT devices. Unlike traditional blockchain networks, 
the device costs a similar amount of energy to communicate compared 
to computing in the wireless blockchain. First, we analyze the energy consumption of maintaining
the blockchain considering the hash power $h$, transmission power $p$, 
and connectivity $c$. Centrality is a metric to measure the influence of 
information propagation. In this paper, we use degree centrality to evaluate 
the connectivity, which reflects the number of the device's neighbors. 
The hash power determines the energy consumption
of computing, and the energy consumption of communication is related
to the connectivity $c$ and transmission power $p$. The total energy
consumption of proposing a block is given as follows:

\begin{equation}
E_{b}=h\tau+\frac{a}{r}pc,\label{eq:Eb}
\end{equation}
where $\tau$ is the average block interval, $a$ is the block size,
and $r$ is the transmission rate. The first term denotes the energy
consumption for solving the hash puzzle, while the second denotes the 
energy consumption for propagating the block.

Next, we discuss the confirmation probability $G$ of IoT devices.
As shown in Fig. 3, the hash power $h$, transmission power $p$, and 
connectivity $c$ affect the block confirmation in different aspects.
Higher hash power brings a faster mining rate and increases the probability 
that its prong becomes the longest legal chain. The transmission power 
determines the Single to Noise Ratio (SNR), which is negatively correlated 
with the communication outage probability. The connectivity reflects the 
number of other IoT devices connected to the device and influences the block 
propagation delay.

\begin{figure}[t]
  \begin{centering}
  \includegraphics[scale=0.33]{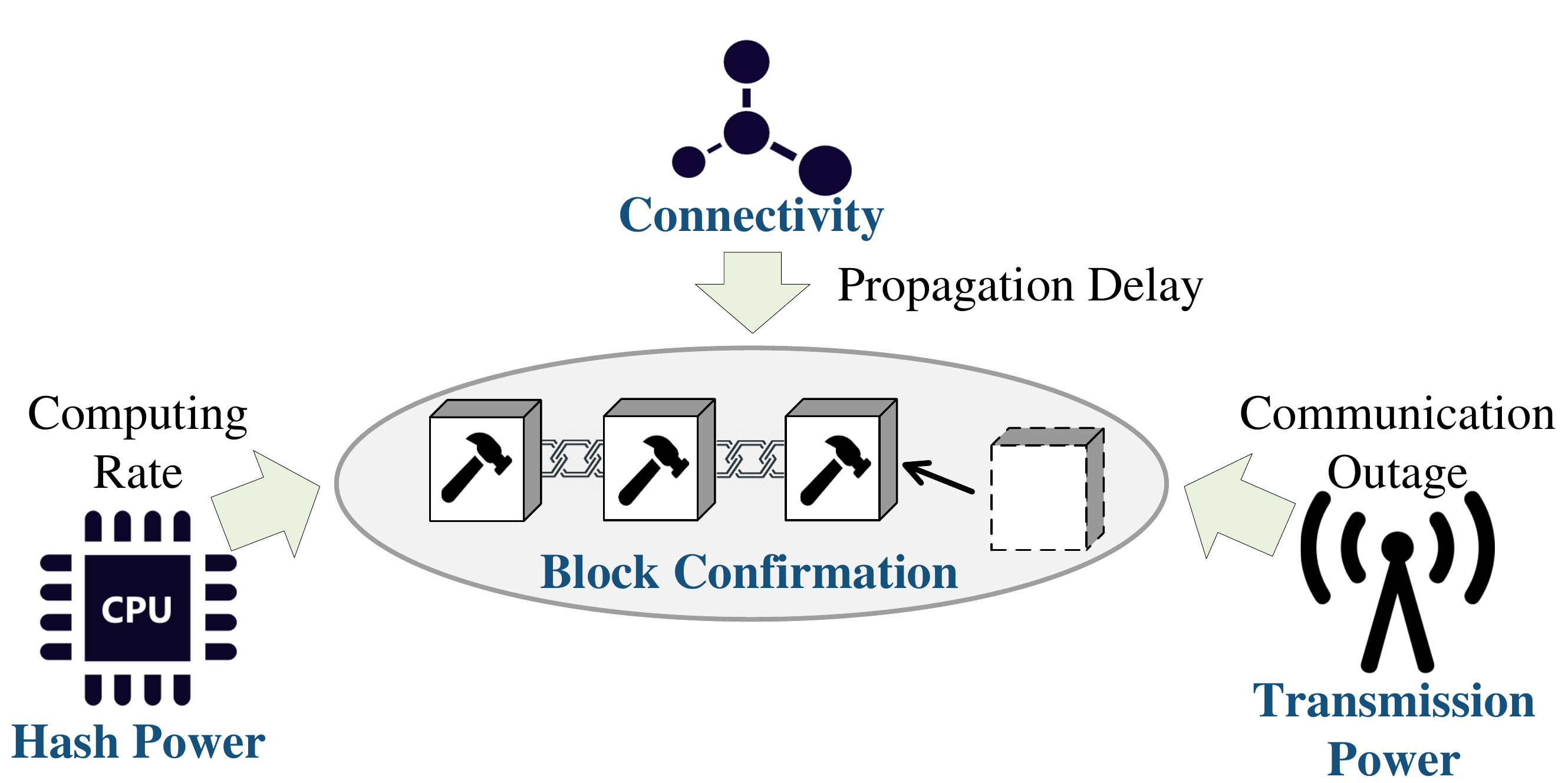}
  \par\end{centering}
  \caption{The impact of IoT device's capabilities on the block confirmation in the wireless blockchain network.}
  \end{figure}

When a forking occurs in the wireless blockchain network, we assume that 
half of IoT devices in the network accept block $i$ (i.e., the block proposed 
by device $i$), and the rest accept block $j$. We define $H _{-i,j}$ is the 
gross hash power of the network except for devices $i$ and $j$. Thus, the hash power 
of the prong $i$ is $H _{-i,j}/2 + h_{i}$ and the hash power of prong $j$ 
is $H _{-i,j}/2 + h_{j}$. The confirmation probability $P_{h}$ of block $i$  
determined by hash power is:

\begin{align}
{{\rm{P}}_h} = \frac{{{H_{ - i,j}}/2 + {h_i}}}{H} = \frac{{0.5(Z - 2)\overline{h}  + h_{i}}}{{Z\overline{h}}},\
\end{align}
where $Z$ is the network size, and $\overline{h}$ is the hash power expectation 
of the IoT device. Since the IoT device cannot know the hash power of 
others, we use the hash power expectation to estimate. In the IoT network, 
the hash power has little difference among devices and slightly affects the 
block confirmation, especially when the network size $Z$ is large. 

Communication outage is universal in the wireless network due
to the randomness and uncertainty of wireless channels. Here, we just 
consider the outage event caused by signal fading. We assume that 
the signal fading is exponentially distributed $\mid\alpha\mid^{2}$ 
with parameter $\sigma^{2}.$ Inspired by \cite{Laneman2004}, we derive 
the outage probability as follows:

\begin{align}
P_{out}= \Pr\{\mid\alpha\mid^{2}<\frac{2^{\delta}-1}{\mathrm{\rho }}\} = 1-\exp(-\frac{2^{\delta}-1}{\mathrm{\rho }\sigma^{2}}),\label{eq:tp_cp1}
\end{align}

\begin{equation}
\delta=\frac{r}{B},\label{eq:tp_cp2}
\end{equation}

\begin{equation}
\mathrm{\rho }=\frac{p}{N_{0}},\label{eq:tp_cp3}
\end{equation}
where $B$ is the bandwidth, $\rho$ is the SNR, $\delta$ is the spectral efficiency and
$N_{0}$ is noise power. From equations \eqref{eq:tp_cp1}, \eqref{eq:tp_cp2},
and \eqref{eq:tp_cp3}, we find that the transmission power $p$ determines
the communication outage probability during the block propagation process.

The impact of connectivity on confirmation probability is through influencing 
the block propagation delay. As we mentioned before, we use 
degree centrality to evaluate the connectivity $c$, which reflects the 
number of neighbors of the IoT device. However, the degree centrality is 
related to the network size $Z$ and the average link probability $P_{l}$. For 
example, the same degree centrality has different effects on the confirmation 
probability under different network sizes or average link probability. 
Therefore, it is difficult and unrealistic to model the relationship 
between the connectivity and confirmation probability without considering 
the network size and the averaged link probability. Given the network size
$Z$ and the average link probability $P_{l}$, the confirmation probability 
$P_{c}$ determined by connectivity $c$ can be represented as follows:

\begin{equation}
P_{c}=\beta_{1}-\beta_{2}\ln(c+\beta_{3}),\label{eq:Pc}
\end{equation}
where $\beta_{1}$, $\beta_{2}$, and $\beta_{3}$ are curve-fitting
parameters. The curve-fitting approach is typical in the literature
and has been widely adopted. Note that the function is universal to 
model the connectivity $c$ and confirmation probability $P_{c}$, 
and we just modify the parameters in different network sizes and average 
link probability. We further discuss the combined impact of $c$, $Z$, 
and $P_{l}$ on confirmation probability in Section \mbox{VI}.

Based on equations \eqref{eq:tp_cp1} and \eqref{eq:Pc}, we can derive
the total confirmation probability $G$ as follows:

\begin{equation}
G(h,c,p)=[P_{F}P_{h}+(1-P_{F})]P_{c}(1-P_{out}),\label{eq:G}
\end{equation}
where $P_{F}$ is the probability of occurring the forking. In our paper, the 
forking probability is set to $0.5$.

Then, according to equations \eqref{eq:Eb} and \eqref{eq:G}, we can 
give the utility function $U_{D}$ of the IoT device as:

\begin{equation}
U_{D}=s+G(h,c,p)be-\frac{1}{2}F(h,c,p)e^{2}-\varphi,\label{eq:user utility}
\end{equation}

\begin{equation}
F(h,c,p)=\gamma E_{b},\label{eq:energy cost}
\end{equation}

\begin{equation}
\varphi=\theta s,\label{eq:time cost}
\end{equation}
where $F$ and $\varphi$ denote the cost function of energy and time,
respectively, $\gamma$ is the energy cost coefficient, and $\theta$
is the time cost coefficient. Similar to \cite{Lijing}, we set the
energy cost function as a quadratic function concerning the number of blocks,
which is widely applied in the literature. Equation \eqref{eq:time cost}
represents the time of the IoT device contributing to the blockchain network,
and we consider that time is positively correlated with the reward.
Consequently, we set the time cost as a linear function of salary, and the time  
cost increases with the salary.

\subsection{Wireless Blockchain Model}

From the wireless blockchain perspective, on the one hand, the developer expects
more IoT devices to join the blockchain, where he can benefit from it. On
the other hand, the developer must pay rewards to the devices to incentivize
them, which brings the costs. Therefore, we can define the utility
function of the wireless blockchain as follows:

\begin{align}
U_{B} & =\sum_{k=1}^{K}\sum_{m=1}^{M}\sum_{n=1}^{N}ZQ(h_{k},c_{m},p_{n})\Bigl(\varepsilon G(h_{k},c_{m},p_{n})e_{k,m,n}\nonumber \\
 & \ \ \ -s_{k,m,n}-G(h_{k},c_{m},p_{n})b_{k,m,n}e_{k,m,n}\Bigr),
\end{align}
where $\varepsilon$ is the yield coefficient and is used to evaluate 
the utility brought to the blockchain by devices. While pursuing maximizing 
the wireless blockchain utility, the contract should meet the IR and IC 
conditions to successfully attract and reveal actual information about IoT devices. 
The conditions are given in the following:

\begin{equation}
U_{D}^{k,m,n}(\omega_{k,m,n})\geq0,\label{eq:IR}
\end{equation}

\begin{equation}
U_{D}^{k,m,n}(\omega_{k,m,n})\geq U_{D}^{k,m,n}(\omega^{'}),\forall\omega_{k,m,n}\neq\omega^{'},\label{eq:IC}
\end{equation}
where $\omega$ is the contract item. Equations \eqref{eq:IR} and
\eqref{eq:IC} denote IR and IC conditions respectively.

\subsection{Problem of Adverse Selection}

Usually, the blockchain developer does not know the IoT devices' actual capabilities.
Thus, the developer cannot obtain IoT devices' energy consumption and block
confirmation probability. We define the IoT devices' capabilities are different
over the hash power $h$, transmission power $p$, and connectivity $c$. 
The tuple $(h,c,p)$ denotes 
the IoT device's capability, and the blockchain developer just knows the probability 
distribution $Q(h,c,p)$ of the capability from the past statistical data. We use the device's
capability as its type $(h_{k},c_{m},p_{n}),1\leq k\leq K,1\leq m\leq M,1\leq n\leq N$.
The device's capability belongs to $L$ different types, where $L = K*M*N$.
The IoT devices consume energy and time to maintain the blockchain, and the developer 
offers a contract $(e,R)$ to compensate the IoT devices.

This is a typical \emph{adverse selection }problem. The contract $(e,R)$ 
must meet the IR and IC conditions to reveal the actual capabilities
of IoT devices to overcome the problem. 

\subsection{Problem of Moral Hazard}

As mentioned in Section \mbox{III} {A}, the block may not be confirmed
on the chain due to the forking. Even if the device signs a contract $(e,R)$, 
the corresponding quality of the task completion may not be ensured because of 
the failure of the block confirmation, which causes the problem of 
\emph{moral hazard}. Consequently, the developer must consider 
the quality of the tasks while distributing rewards. Based on the longest 
legal chain principle, it is easy to verify whether the block is 
on the chain and then give rewards to the IoT devices.

\subsection{Problem Formulation on Maximizing Wireless Blockchain Utility} 

According to the above analyses, the wireless blockchain utility maximization
problem can be formulated as follows:

\begin{subequations}
\begin{align}
\max U_{B} & =\sum_{k=1}^{K}\sum_{m=1}^{M}\sum_{n=1}^{N}ZQ(\omega_{k,m,n})\Bigl(\varepsilon G(\omega_{k,m,n})e_{k,m,n}\nonumber \\
 & \ \ \ -s_{k,m,n}-G(\omega_{k,m,n})b_{k,m,n}e_{k,m,n}\Bigr),\label{eq:problem}
\end{align}

\begin{align}
s.t. & \ \ U_{D}^{k,m,n}(\omega_{k,m,n})\geq0,\label{eq:cons1}\\
 & \ \ U_{D}^{k,m,n}(\omega_{k,m,n})\geq U_{D}^{k,m,n}(\omega^{'}),\forall\omega_{k,m,n}\neq\omega^{'}.\label{eq:cons2}
\end{align}
\end{subequations}

The problem \eqref{eq:problem} with constraints \eqref{eq:cons1}
and \eqref{eq:cons2} is difficult to solve directly. We first
resolve the optimal number of blocks $e^{*}$ for the IoT device. Based
on equations \eqref{eq:Eb}, \eqref{eq:user utility}, \eqref{eq:energy cost},
and \eqref{eq:time cost}, it is easy to derive that equation \eqref{eq:user utility}
is a concave function of $e$. We take the first derivative of the
IoT device's utility for the number of blocks $e$ and set it to zero:

\begin{equation}
\frac{\partial U_{D}(\omega_{k,m,n})}{\partial e}=0.\label{eq:first dericative}
\end{equation}

Then, the optimal number of blocks $e^{*}$ can be obtained according to equation
\eqref{eq:first dericative} as follows:

\begin{equation}
e_{k,m,n}^{*}=\frac{G(\omega_{k,m,n})b_{k,m,n}}{F(\omega_{k,m,n})}.\label{eq:optimal e}
\end{equation}

From equation \eqref{eq:optimal e}, we find that the bonus can represent
the number of blocks. By substituting $e_{k,m,n}^{*}$ into equation
\eqref{eq:user utility}, we can rewrite the IoT device's utility
function as:

\begin{equation}
U_{D}^{k,m,n}(\omega_{k,m,n})=\frac{G^{2}(\omega_{k,m,n})b_{k,m,n}^{2}}{2F(\omega_{k,m,n})}-(\theta-1)s_{k,m,n}.\label{eq:user utility1}
\end{equation}

Similarly, we substitute the optimal number of blocks $e^{*}$ into equation
\eqref{eq:problem}, and the problem is rewritten as follows:

\begin{subequations}
\begin{align}
\max_{(s,b)}U_{B} & =\sum_{k=1}^{K}\sum_{m=1}^{M}\sum_{n=1}^{N}ZQ(\omega_{k,m,n})\Bigl(\frac{\varepsilon G^{2}(\omega_{k,m,n})b_{k,m,n}}{F(\omega_{k,m,n})}\nonumber \\
 & \ \ \ -s_{k,m,n}-\frac{G^{2}(\omega_{k,m,n})b_{k,m,n}^{2}}{F(\omega_{k,m,n})}\Bigr),\label{eq:problem-1}
\end{align}

\begin{align}
s.t. & \ \ U_{D}^{k,m,n}(\omega_{k,m,n})\geq0,\label{eq:cons1-1}\\
 & \ \ U_{D}^{k,m,n}(\omega_{k,m,n})\geq U_{D}^{k,m,n}(\omega^{'}),\forall\omega_{k,m,n}\neq\omega^{'}.\label{eq:cons2-1}
\end{align}
\end{subequations}

To solve the problem of maximizing the wireless blockchain utility, 
we should design the salary $s$ and bonus $b$ elaborately. In the following, 
we give the process of designing the optimal contract and analyzing the feasibility 
of the contract.

\section{Optimal Contract Design}

We resort to an effective method to convert the multi-dimensional
contract problem into a one-dimensional problem in this selection.
Then, we relax the IR and IC conditions and obtain the optimal contract
($e,R$).

\subsection{Conversion of IoT Device Type}

According to equation \eqref{eq:user utility1}, the IoT device's
type $(h_{k},c_{m},p_{n})$ contains three-dimensional attributes:
hash power $h$, connectivity $c$, and transmission power $p$, which 
determine the device's energy cost and block confirmation probability.  
Directly resolving the problem considering three attributes is complicated.
Here, we introduce the preference order $\lambda$ to represent the
IoT device's new type. From the analyses in Section \mbox{IV}, the
device's capability not only affects the energy consumption but also
the confirmation probability. Therefore, we define the new IoT device's
type $\lambda$ as follows:

\begin{equation}
  \lambda=\frac{G^{2}(h,c,p)}{2F(h,c,p)}.\label{eq:new type}
\end{equation}

\begin{figure}[t]
  \subfloat[Connectivity $c=1$]{\includegraphics[scale=0.18]{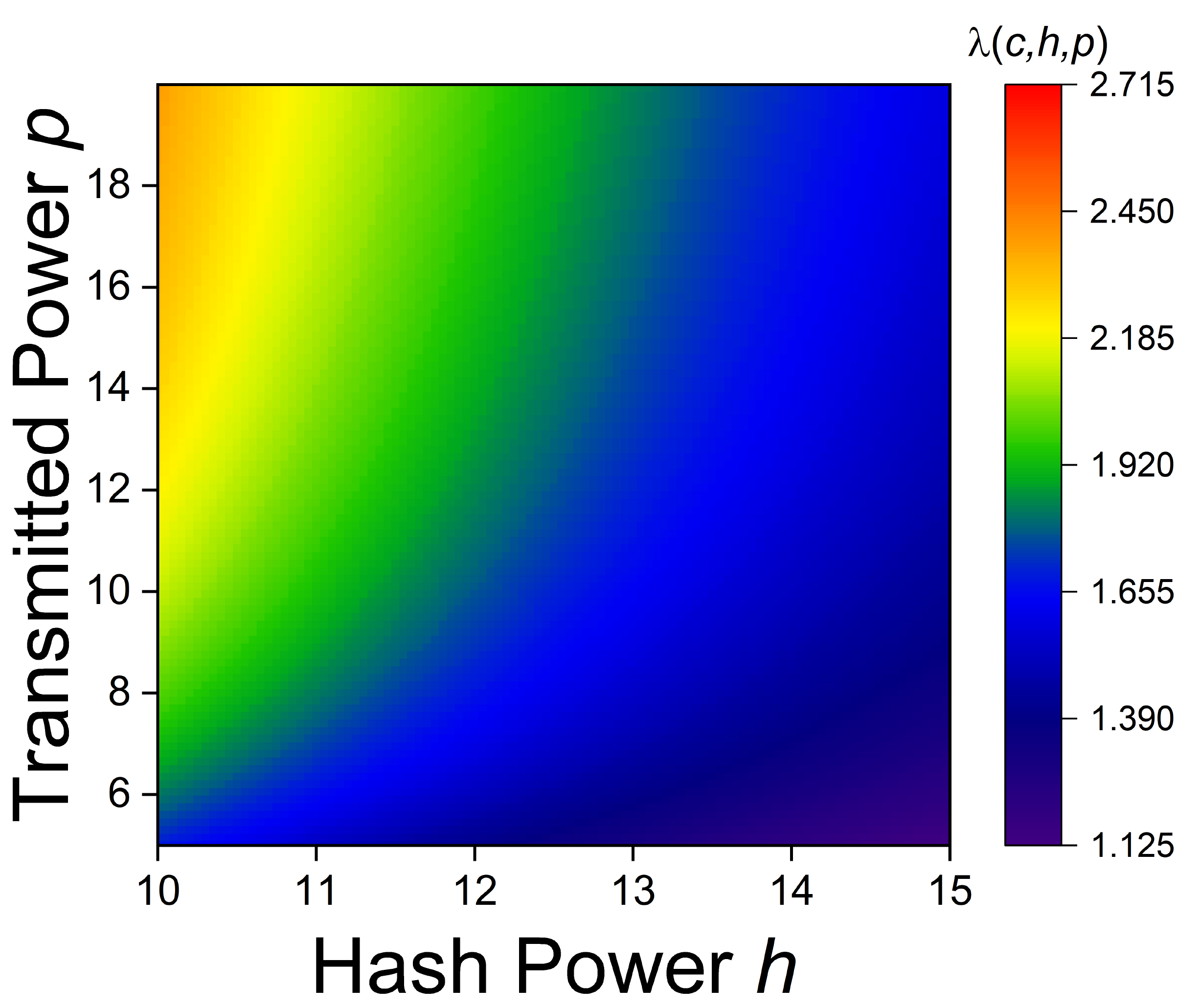}
  
  }\subfloat[Connectivity $c=10$]{\includegraphics[scale=0.18]{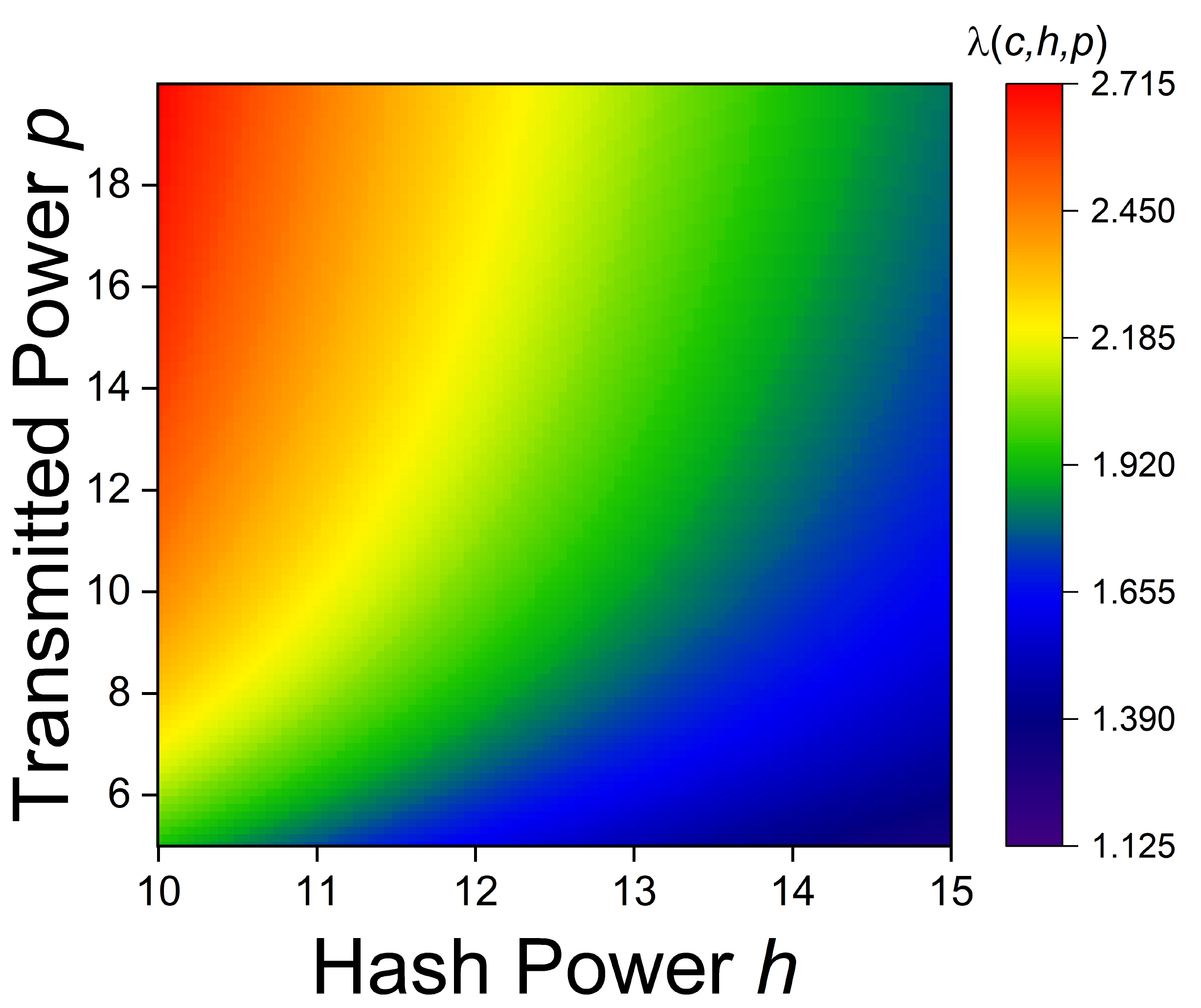}
  
  }
  
  \subfloat[Connectivity $c=15$]{\includegraphics[scale=0.18]{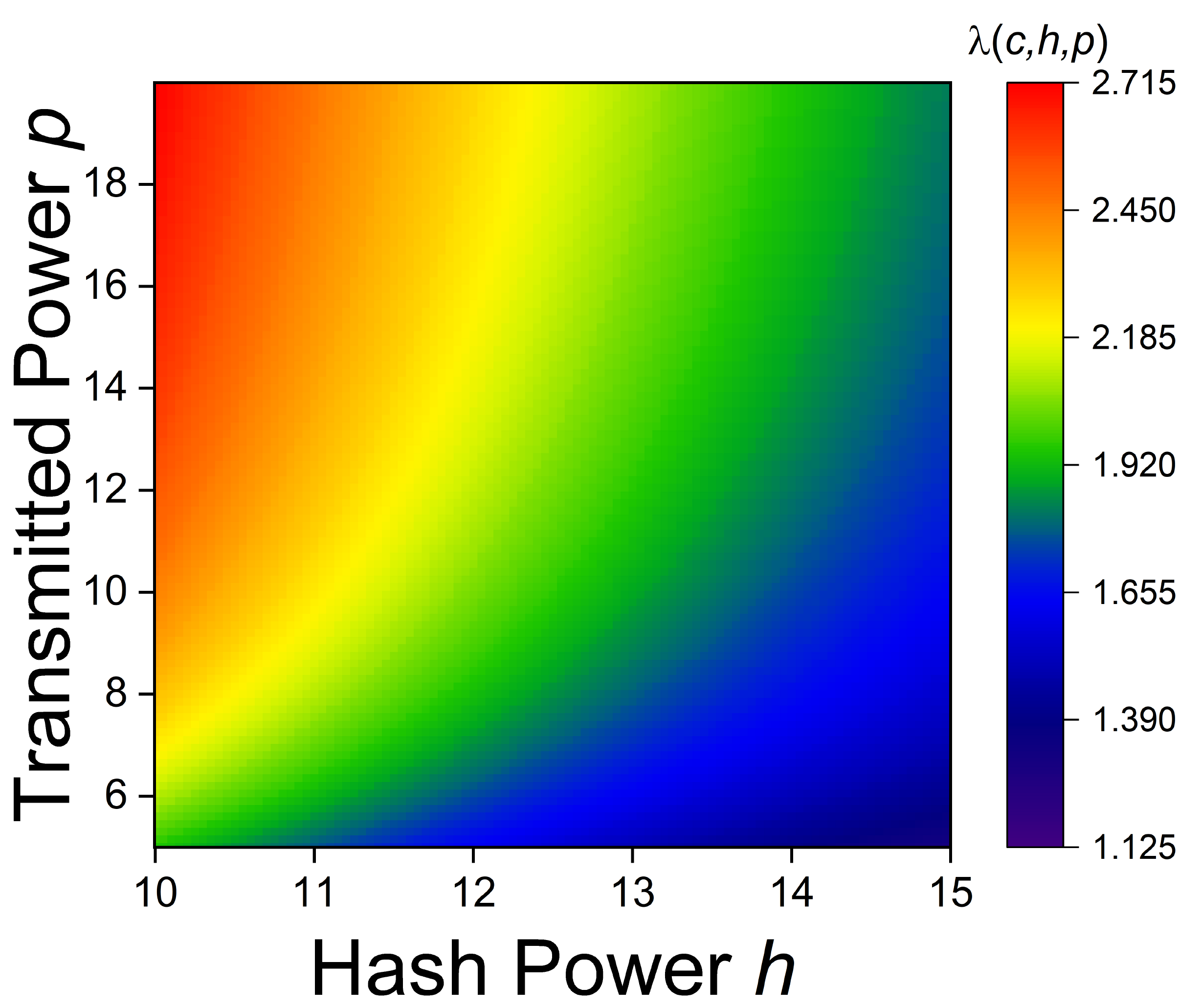}
  
  }\subfloat[Connectivity $c=25$]{\includegraphics[scale=0.18]{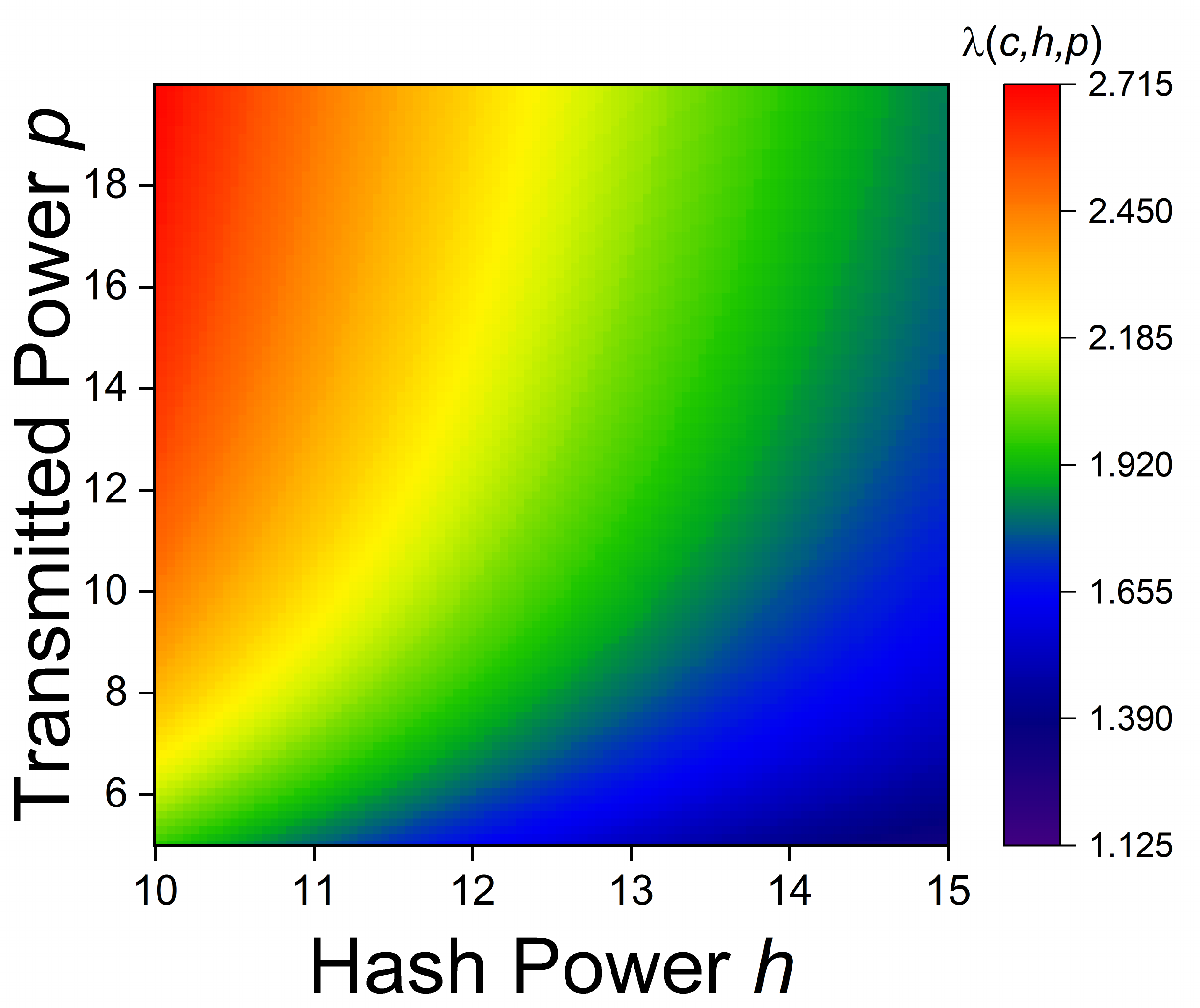}
  
  }\caption{The IoT device's preference order.}
\end{figure}

The device's preference order is illustrated in Fig. 4. The color represents the
value of the preference order $\lambda$. The closer the color is to red, the higher 
the value, and vice versa. According to the preference order, we can quickly 
redefine the types of devices. The more excellent value of $\lambda$ means
the IoT device has a higher confirmation probability and lower energy
consumption, which is the preferred type of wireless blockchain developer.
We reclassify the IoT device's type based on equation \eqref{eq:new type}.
The $L$ types are sorted in non-descending order in the
following:

\begin{equation}
\lambda_{1}\leq\ldots\leq\lambda_{i}\leq\ldots\leq\lambda_{L}.
\end{equation}

Therefore, the IoT device's utility is represented as:

\begin{equation}
U_{D}^{i}(\omega_{i})=\lambda_{i}b_{i}^{2}-(\theta-1)s_{i}.\label{eq:device utility}
\end{equation}

Intuitively, the higher type $\lambda_{i}$ can afford a great number of
blocks and obtain more rewards $R.$ To this end, the contract $(e_{i},R_{i})$
should be in accordance with the type $\lambda_{i}.$ The salary $s$
and bonus $b$ satisfy the following monotonicity constraints: 

\begin{align}
s_{1} & \leq\ldots\leq s_{i}\leq\ldots\leq s_{L},\nonumber \\
b_{1} & \leq\ldots\leq b_{i}\leq\ldots\leq b_{L}.\label{eq:monotonous cons}
\end{align}

Monotonicity constraints guarantee the correspondence between
contract and reward, which is an essential principle in contract theory.

By introducing the preference order $\lambda$, we successfully convert the multi-dimensional
contract into a one-dimensional contract. The post-converted
problem is rewritten as follows:

\begin{subequations}
\begin{align}
\max_{(s_{i},b_{i})}U_{B}= & \sum_{i=1}^{L}ZQ(\lambda_{i})\Bigl(2\varepsilon\lambda_{i}b_{i}-s_{i}-2\lambda_{i}b_{i}^{2}\Bigr),\label{eq:one-dimensional contract}\\
s.t. & \ \ U_{D}^{i}(\omega_{i})\geq0,1\leq i\leq L,\label{eq:one IR}\\
 & \ \ U_{D}^{i}(\omega_{i})\geq U_{D}^{i}(\omega_{j}),\forall i\neq j,\label{eq:one IC}\\
 & {\textstyle \ \ {\displaystyle Constraints\ in}\ }\eqref{eq:monotonous cons}.\nonumber 
\end{align}
\end{subequations}

\subsection{Constraints Reduction}

Although the problem has been converted into a one-dimensional
contract, it is still difficult to resolve due to excessive constraints.
From equations \eqref{eq:one-dimensional contract}, \eqref{eq:one IR},
and \eqref{eq:one IC}, the problem has $\mathit{L}$ IR constraints
and $\mathit{L(\mathit{L}-\mathrm{1})}$ IC constraints. All of
these constraints are non-convex and not straightforward to handle.
Thus, we need to reduce the IR and IC constraints. 

First, we relax the IR constraints by introducing Lemma 1.
\begin{lem}
If the $\lambda_{1}$ type of device satisfies the IR constraint,
all types of devices will satisfy the IR constraints.
\end{lem}
\begin{IEEEproof}
According to IC constraints, we have

\begin{equation}
U_{D}^{i}(\omega_{i})\geq U_{D}^{i}(\omega_{1}).\label{eq:proof IR}
\end{equation}Then based on equations \eqref{eq:monotonous cons} and \eqref{eq:proof IR},
we can derive

\begin{equation}
U_{D}^{i}(\omega_{1})\geq U_{D}^{1}(\omega_{1}), 
\end{equation}

\begin{equation}
U_{D}^{i}(\omega_{i})\geq U_{D}^{1}(\omega_{1}).
\end{equation}

Therefore, we conclude that $\forall1\leq i\leq L$, if $U_{D}^{1}(\omega_{1})\geq0$, $U_{D}^{i}(\omega_{i})\geq0$.
The IR constraints are met, and Lemma 1 is proved.
\end{IEEEproof}
Lemma 1 guarantees reducing the $L$ IR constraints to one
constraint. Before taking up reducing the IC constraints, we 
introduce the following concepts \cite{Game}.
\begin{enumerate}
\item \emph{Downward Incentive Constraints (DICs).} The IC constraints between
type $\lambda_{i}$ and $\lambda_{j},1\leq j\leq i-1$ are called
DICs.
\item \emph{Local Downward Incentive Constraints (LDIC).} The IC constraint
between type $\lambda_{i}$ and $\lambda_{j},j=i-1$ is called LDIC.
\item \emph{Upward Incentive Constraints (UICs). }The IC constraints between
type $\lambda_{i}$ and $\lambda_{j},i+1\leq j\leq L$ are called
UICs.
\item \emph{Local Upward Incentive Constraint (LUIC).} The IC constraint
between type $\lambda_{i}$ and $\lambda_{j},j=i+1$ is called LUIC.
\end{enumerate}
Obviously, the IC constraints are composed of DICs and UICs. By virtue
of the above concepts, we reduce the IC constraints by introducing
Lemma 2.
\begin{lem}
For any IoT device's type $\lambda_{i}$, if the LDIC holds, then
all DICs hold, and the same with LUIC and UICs.
\end{lem}
\begin{IEEEproof}
Based on LDIC, we have

\begin{align}
U_{D}^{i}(\omega_{i}) & \geq U_{D}^{i}(\omega_{i-1}),\label{eq:lemma 2}\\
U_{D}^{i-1}(\omega_{i-1}) & \geq U_{D}^{i-1}(\omega_{i-2}).\label{eq:lemma 2 1}
\end{align}

Then we define the following variates $\delta_{1}$ and $\delta_{2}$
to reduce the IC constraints:

\begin{align}
\delta_{1}= & (\lambda_{i}-\lambda_{i-1})b_{i-1}^{2},\label{eq:lemma 2 2}\\
\delta_{2}= & (\lambda_{i}-\lambda_{i-1})b_{i-2}^{2}.\label{eq:lemma 2 3}
\end{align}

According to the monotonicity constraints \eqref{eq:monotonous cons},
it is easy to derive that $\delta_{1}\geq\delta_{2}$. We add equations
\eqref{eq:lemma 2 2} and \eqref{eq:lemma 2 3} to the left and right sides of 
the inequality \eqref{eq:lemma 2 1}, respectively.
Then, we can have:

\begin{equation}
U_{D}^{i}(\omega_{i-1})\geq U_{D}^{i}(\omega_{i-2}).\label{eq:lemma 2 4}
\end{equation}

With equations \eqref{eq:lemma 2} and \eqref{eq:lemma 2 4},
we obtain

\begin{equation}
U_{D}^{i}(\omega_{i})\geq U_{D}^{i}(\omega_{i-2}).\label{eq:lemma 2 5}
\end{equation}

Iterate steps \eqref{eq:lemma 2}, \eqref{eq:lemma 2 1}, \eqref{eq:lemma 2 4},
and \eqref{eq:lemma 2 5}, we derive the following simultaneous inequalities:

\begin{align}
U_{D}^{i}(\omega_{i})\geq  U_{D}^{i}(\omega_{i-2})
\geq  U_{D}^{i}(\omega_{i-3})
\geq \ldots
\geq  U_{D}^{i}(\omega_{1}).
\end{align}

As a consequence, for the type $\lambda_{i}$ of the IoT device, if the
contract satisfies the LDIC, it satisfies the DICs. Similarly, we
can prove the LUIC and UICs by the above steps. Lemma 2 is proven.
\end{IEEEproof}
Using Lemma 2, we reduce $\mathit{L(L-\mathrm{1})}$
IC constraints to $\mathit{L}$ constraints. We redefine
the problem with the LDIC and LUIC as follows:

\begin{subequations}
\begin{align}
\max_{(s_{i},b_{i})}U_{B}= & \sum_{i=1}^{L}ZQ(\lambda_{i})\Bigl(2\varepsilon \lambda_{i}b_{i}-s_{i}-2\lambda_{i}b_{i}^{2}\Bigr),\label{eq:one-dimensional contract-1}\\
s.t. & \ \ U_{D}^{1}(\omega_{1})=0,\label{eq:one IR-1}\\
 & \ \ U_{D}^{i}(\omega_{i})=U_{D}^{i}(\omega_{i-1}),1\leq i\leq L,\label{eq:one IC-1}\\
 & {\textstyle \ \ {\displaystyle Constraints\ in}\ }\eqref{eq:monotonous cons}.\nonumber 
\end{align}
\end{subequations}

\subsection{Salary and Bonus Design}

To resolve the problem \eqref{eq:one-dimensional contract-1}, we
need to derive the salary $s$ and bonus $b$. We first give the salary
$s_{i}$ with respect to type $\lambda_{i}$. Then we replace the
$s_{i}$ and calculate the optimal bonus $b_{i}^{*}$. The salary
is obtained by Theorem 1 as follows:

\begin{thm}
For any device's type $\lambda_{i}$, if the contract is feasible,
the salary satisfies

\begin{equation}
s_{i}^{*}=\sum_{t=1}^{i}\frac{\lambda_{t}}{\theta-1}(b_{t}^{2}-b_{t-1}^{2}),1\leq i\leq L.\label{eq:salary}
\end{equation}
\end{thm}
\begin{IEEEproof}
According to equation \eqref{eq:one IR-1}, we can calculate the salary
of type $\lambda_{1}$

\begin{equation}
s_{1}=\frac{\lambda_{1}}{\theta-1}b_{1}^{2}.\label{eq:s1}
\end{equation}

Then based on equations \eqref{eq:one IC-1} and \eqref{eq:s1}, we
further derive

\begin{align}
s_{2} & =\frac{\lambda_{2}}{\theta-1}(b_{2}^{2}-b_{1}^{2})+s_{1}\nonumber \\
 & =\frac{\lambda_{2}}{\theta-1}(b_{2}^{2}-b_{1}^{2})+\frac{\lambda_{1}}{\theta-1}b_{1}^{2}.\label{eq:s2}
\end{align}

Repeat steps \eqref{eq:s1} and \eqref{eq:s2}, we can obtain

\begin{align}
s_{i} & =\frac{\lambda_{i}}{\theta-1}(b_{i}^{2}-b_{i-1}^{2})+s_{i-1}\nonumber \\
 & =\frac{\lambda_{i}}{\theta-1}(b_{i}^{2}-b_{i-1}^{2})+\ldots+\frac{\lambda_{1}}{\theta-1}b_{1}^{2}\nonumber \\
 & =\sum_{t=1}^{i}\frac{\lambda_{t}}{\theta-1}(b_{t}^{2}-b_{t-1}^{2}).
\end{align}

Therefore, Theorem 1 is proved.
\end{IEEEproof}
From equation \eqref{eq:salary}, we find that the salary $s$
can be expressed by bonus $b$. By replacing $s$ with $b$, we derive
the final form of the problem in the following:

\begin{align}
\max_{b_{i}}U_{B}= & \sum_{i=1}^{L}Z\Bigl(Q(\lambda_{i})\bigl(2\varepsilon\lambda_{i}b_{i}-2\lambda_{i}b_{i}^{2}\bigr)\nonumber \\
 & -\frac{\lambda_{i}}{\theta-1}\bigl(b_{i}^{2}-b_{i-1}^{2}\bigr)\sum_{t=i}^{L}Q(\lambda_{t})\Bigr),\label{eq:final problem}\\
s.t. & {\textstyle \ \ {\displaystyle Constraints \ in}\ }\eqref{eq:monotonous cons}.\nonumber 
\end{align}

Obviously, equation \eqref{eq:final problem} is a concave function
with respect to $b_{i}$. We can derive the optimal $b_{i}^{*}$ by
calculating $\frac{\partial U_{B}}{\partial b_{i}}=0$ without considering
the constraints \eqref{eq:monotonous cons}. However, the results
may not satisfy monotonicity. Inspired by \cite{Gao2011} and
\cite{Xiong2020}, we adopt an Ironing Algorithm to adjust the results.
The algorithm is presented in Algorithm 1. After obtaining the optimal
$b_{i}^{*}$, we can calculate the corresponding $s_{i}$ and $e_{i}$
according to equations \eqref{eq:salary} and \eqref{eq:optimal e}.

\begin{algorithm}[t]
  \textbf{Input : }The bonus sequence $\tilde{\mathbf{b}}=\left\{ b_{1},\ldots b_{i},\ldots b_{L}\right\} $.
  
  \textbf{Output: }The monotonous bonus sequence $\mathbf{b}^{*}$.
  
  \hspace{5bp}1:\textbf{ begin}
  
  \hspace{5bp}2: \ \ \textbf{\ \ while} $\tilde{\mathbf{b}}$ violates
  the monotonicity constraints \textbf{do}
  
  \hspace{5bp}3: \ \ \ \ \ \ \ \ Find an infeasible sub-sequence
  $\mathbf{\hat{b}=}\left\{ b_{i},\ldots b_{i+n}\right\} $
  
  \hspace{5bp}4: \ \ \ \ \ \ \ \ $\tilde{\mathbf{b}}\leftarrow\tilde{\mathbf{b}}\setminus\hat{\mathbf{b}}$
  
  \hspace{5bp}5: \ \ \ \ \ \ \ \ $b_{j}=\textrm{arg}\max_{b}\sum_{j=i}^{i+n}U_{B},i\leq j\leq i+n$
  
  \hspace{5bp}6: \ \ \ \ \ \ \ \ $\mathbf{b^{*}}\leftarrow\tilde{\mathbf{b}}\cup\hat{\mathbf{b}}$
  
  \hspace{5bp}7: \ \ \ \ \ \ \ \ Sort bonus $\mathbf{b^{*}}$
  in ascending order by subscript $i$
  
  \hspace{5bp}8: \ \ \ \ \textbf{end while}
  
  \hspace{5bp}9: \textbf{end}
  
  \caption{Ironing Algorithm}
  \end{algorithm}

\section{Performance Evaluation}

In this section, we present the performance of our proposed contract
and the impact of various parameters on the utility of the IoT device
and blockchain network. We consider that there are $Z$ IoT devices
in the network, which belong to $\mathit{L}$ different types. We
set the IoT device's type subject to a uniform distribution. To
be concrete, $h\sim U  (10,15)$, $c\sim U(1,20)$,
$p\sim U(5,20)$. Based on the clustering method and equation
\eqref{eq:new type}, we partition the capability into $48$ types. The blockchain 
parameters are in reference to bitcoin \cite{Nakamoto2017}. Table 2 lists 
the default parameters. Not otherwise specified, the parameter is set as the default value.

\begin{table}[tbh]
    \caption{Default Parameter Values}
    \centering{}%
    \begin{tabular}{lclc}
    \hline 
    Parameters & Values & Parameters & Values\tabularnewline
    \hline 
    $N$ & 100 & $A$ & 1000\tabularnewline
    $B$ & 2$\times10^{6}$ & $N_{0}$ & 3.98$\times10^{-3}$\tabularnewline
    $\tau$ & 600 & $r$ & 2000\tabularnewline
    $P_{l}$ & 0.2 & $\varepsilon$ & 400\tabularnewline
    $\theta$ & 1.5 & $\gamma$ & 1$\times10^{-4}$\tabularnewline
    \hline 
    \end{tabular}
\end{table}

\subsection{Evaluation of the Probability $P_{c}$}

As we mentioned in Section \mbox{III}, the function of $P_{c}$ is
given according to equation \eqref{eq:Pc}. We set $\beta_{1}=0.97575$,
$\beta_{2}=0.03006$, and $\beta_{3}=0.00411$, and Fig. 5 compares
the fitting values and original values. To better utilize the function in
reality, we regard the normalized degree centrality as
connectivity. In Fig. 5, we derive that the Adjusted R-Square of the
fitting curve is $0.97421$ and the RMSE is $0.00517$, which demonstrates
the effectiveness of the fitting. With the increase of connectivity, 
the probability $P_{c}$ is on the rise and asymptotic to 1. It is intuitive because the extensive
connectivity reduces the propagation delay, which brings the advantage of winning 
the block competition. Fig. 5 demonstrates that communication advantage helps devices 
to win the competition of block property. 
Higher connectivity makes more devices receive the block simultaneously, which is more prominent in 
wireless IoT networks.

Then we discuss the impact of the network size $z$ and average link
probability $P_{l}$. Fig. 6 (a) illustrates how the link
probability affects the probability $P_{c}$
while the network size is fixed at $100$. In
the mass, the probability $P_{c}$ is higher under a larger average
link probability $P_{l}$. The difference in probability $P_{c}$
is slight among different connectivity under a higher average link probability $P_{l}$
network. It is because the higher link probability $P_{l}$ increases
the number of average links and reduces the propagation delay of
the entire network. Moreover, there is an interesting phenomenon. 
For example, while the connectivity $c$ exceeds 0.3, the same connectivity $c$ has a 
higher probability $P_{c}$ at $P_{l}=0.3$ than $P_{l}=0.4$.
This phenomenon seems counterintuitive because the same
connectivity has a different influence in networks with different average 
link probabilities. Though devices are with the same connectivity, they have 
different influences in networks with different average
link probabilities. The device with 0.35 connectivity may have the most communication links 
at $P_{l}=0.3$ but is the common connectivity at $P_{l}=0.4$. Thus, improving the connectivity 
can significantly increase the confirmation probability in the network with sparse connections. 
From Fig. 6 (a), the higher link probability
and more extensive connectivity both raise the probability $P_{c}$
by reducing the delay of the whole network. Nevertheless, the improvement
is asymptotic to a stable value, especially when $P_{l}$ or $c$ is large.

\begin{figure}[t]
  \includegraphics[scale=0.38]{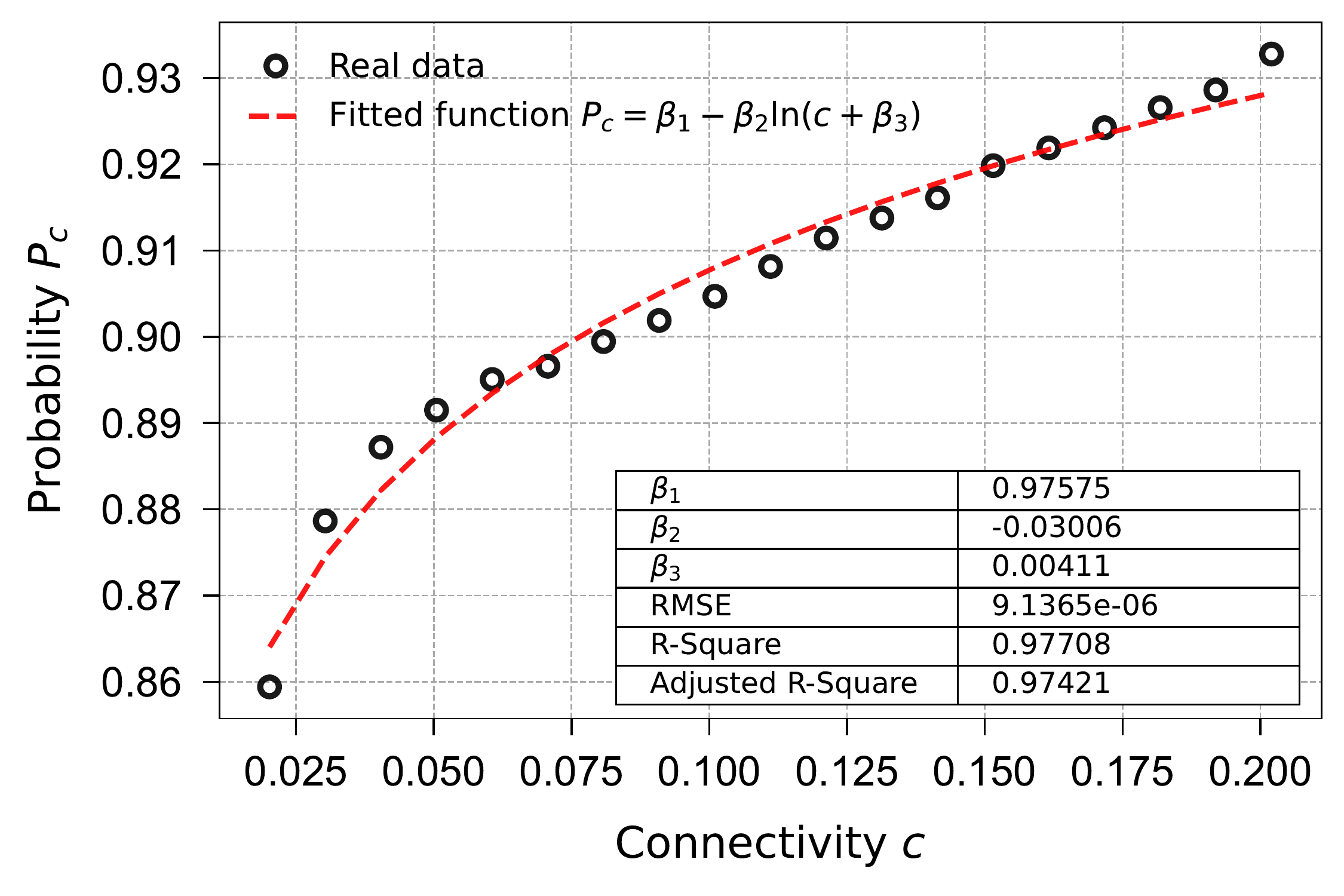}
  \caption{Verification of the confirmation probability function (6).}
\end{figure}

\begin{figure}[t]
  \subfloat[Impact of connectivity $c$ on probability $P_{c}$ under different average link probabilities $P_{l}$ when the network size $Z=100$.]{\includegraphics[scale=0.38]{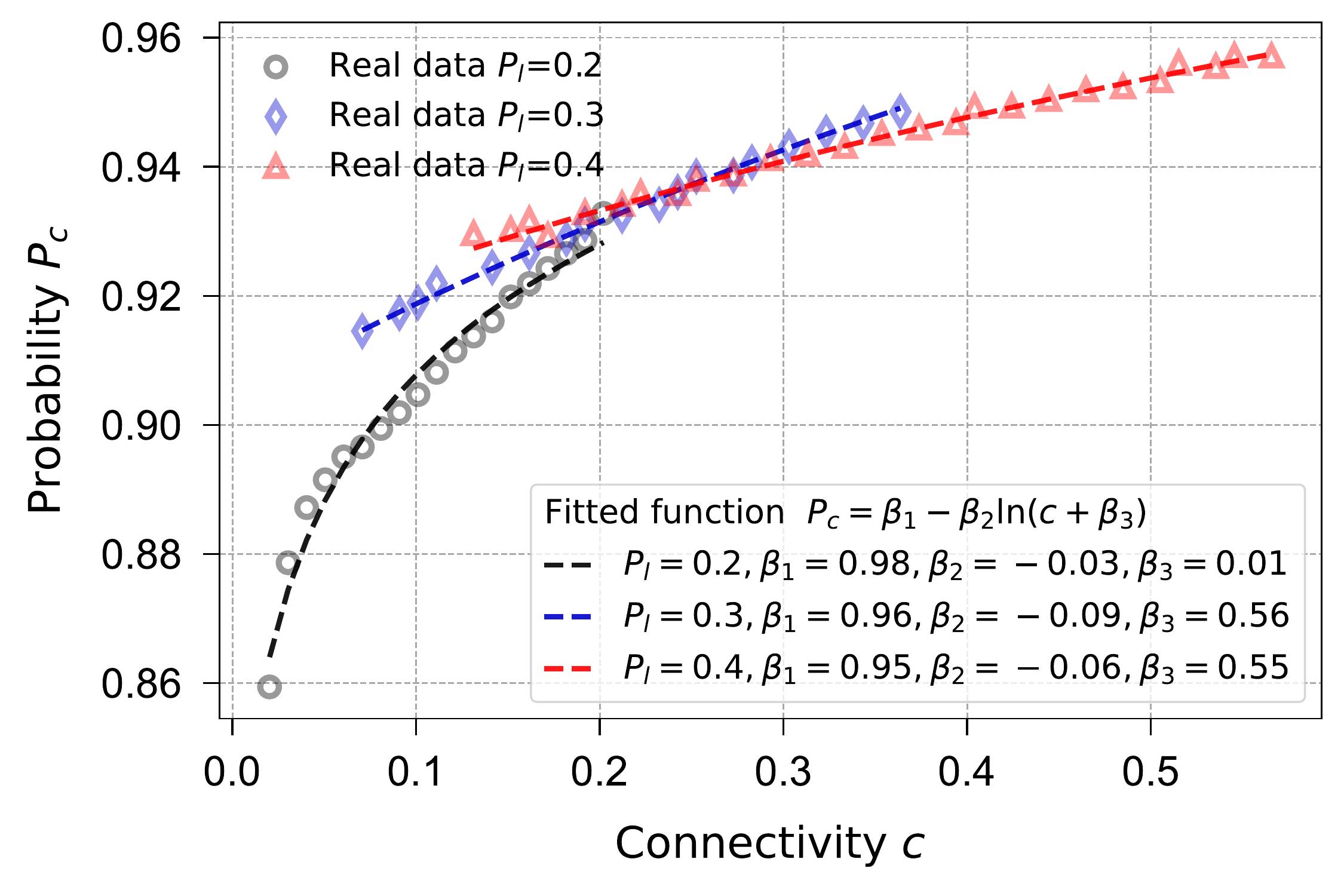}}
  
  \subfloat[Impact of connectivity $c$ on probability $P_{c}$ under different network sizes $Z$ when the average link probability $P_{l}=0.2$.]{\includegraphics[scale=0.38]{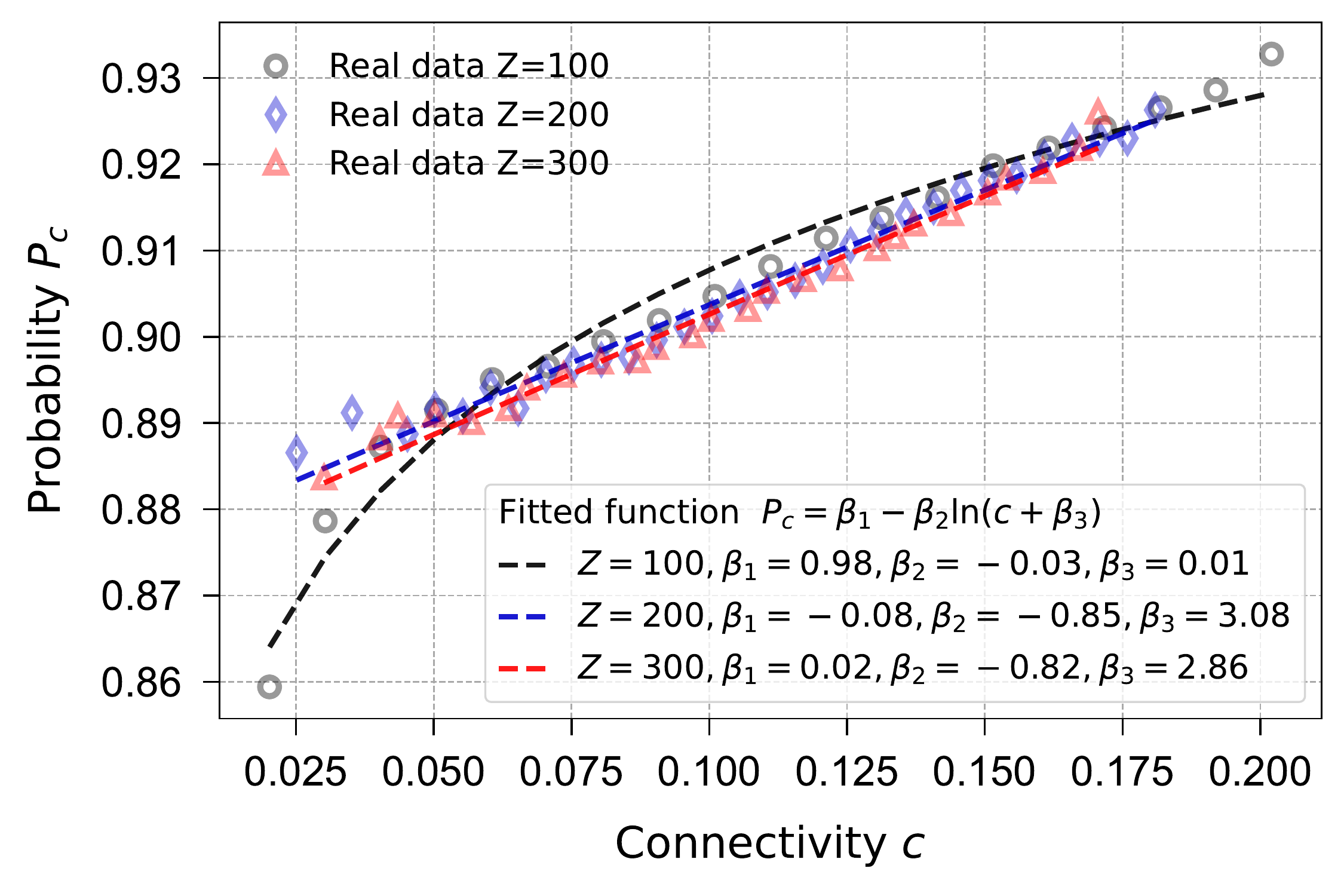}}
  \caption{The impact of the connectivity $c$ on probability $P_{c}$ under different network sizes $Z$ and average link probabilities $P_{l}$,
  where the dot represents the simulation results and the dashed line
  represents the fitted function.}
  \end{figure}

The impact of network size $Z$ is complicated. On the one hand,
the larger network size brings a larger number of average links under
the same average link probability $P_{l}$. On the other hand, the
larger network size also increases communication delay. In Fig.
6 (b), while fixing $P_{l}=0.2$, the probability $P_{c}$ is lower
in the larger network size generally. The expansion in network size
increases the propagation delay for IoT devices, which causes more forking
events. When the network size ranges from 100 to 200, the confirmation probability $P_{c}$ 
becomes higher among small connectivity values. Benefiting from the expansion of the network size, devices 
own larger connections at the same connectivity to improve the propagation of blocks. This is the positive 
influence of network size increase. 
However, when the network size ranges from 200 to 300, a larger network size decreases the confirmation 
probability due to blocks needing more time to spread across the network. Expansion of network size also 
brings negative effects on the confirmation of blocks.
In addition, the IoT device obtains a higher
probability $P_{c}$ in a small-scale network as the connectivity
is large enough. For example, the device with normalized connectivity
of $0.2$ is a core node with a network size $Z=100$. Obviously,
the network size has a two-sided impact on confirmation probability 
compared to the average link probability.

\subsection{Impact of Hash Power $h$, Connectivity $c$, and Transmission
Power $p$}

First, we discuss the impact of the hash power $h$, connectivity
$c$, and transmission power $p$ on the IoT device's energy consumption.
As Fig. 7 shows, the types of IoT devices are divided into $48$ categories,
$h=\left\{ 11,12,13\right\}, c=\left\{ 3,10,15,20\right\}, p=\left\{ 5,10,15,20\right\}$.
From Fig. 7, we observe that the hash power $h$ has a
bigger impact on energy consumption. In contrast, the connectivity
$c$ and the transmission power $p$ slightly affect the energy consumption.
It illustrates that the IoT device consumes more energy for computing
than for communicating. In reality, while participating in the PoW-based
blockchain network, the device mainly consumes energy for computing. 
Moreover, the more capable the device is, the more significant the
difference in energy consumption between them. For example, 
the difference in energy consumption when $p=20$ is more prominent
than $p=5$ while the connectivity $c$ varies from $3$ to $20$. That is why 
the more powerful IoT devices obtain more rewards. 

\begin{figure}[t]
  \includegraphics[scale=0.38]{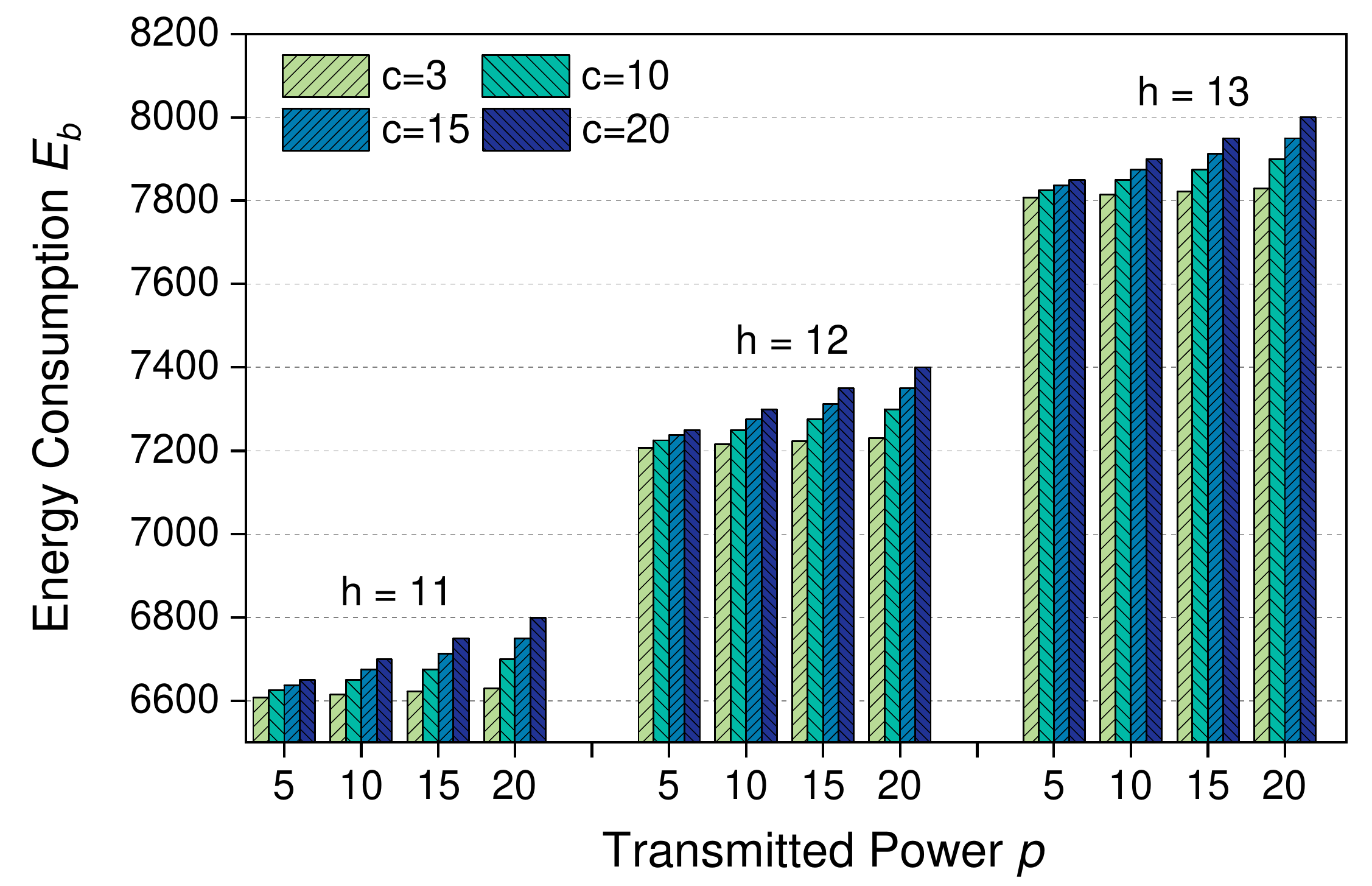}
  \caption{The impact of hash power $h$, connectivity $c$, and transmission
  power $p$ on the IoT device's energy consumption $E_{b}$.}
\end{figure}

\begin{figure}[t]
  \includegraphics[scale=0.38]{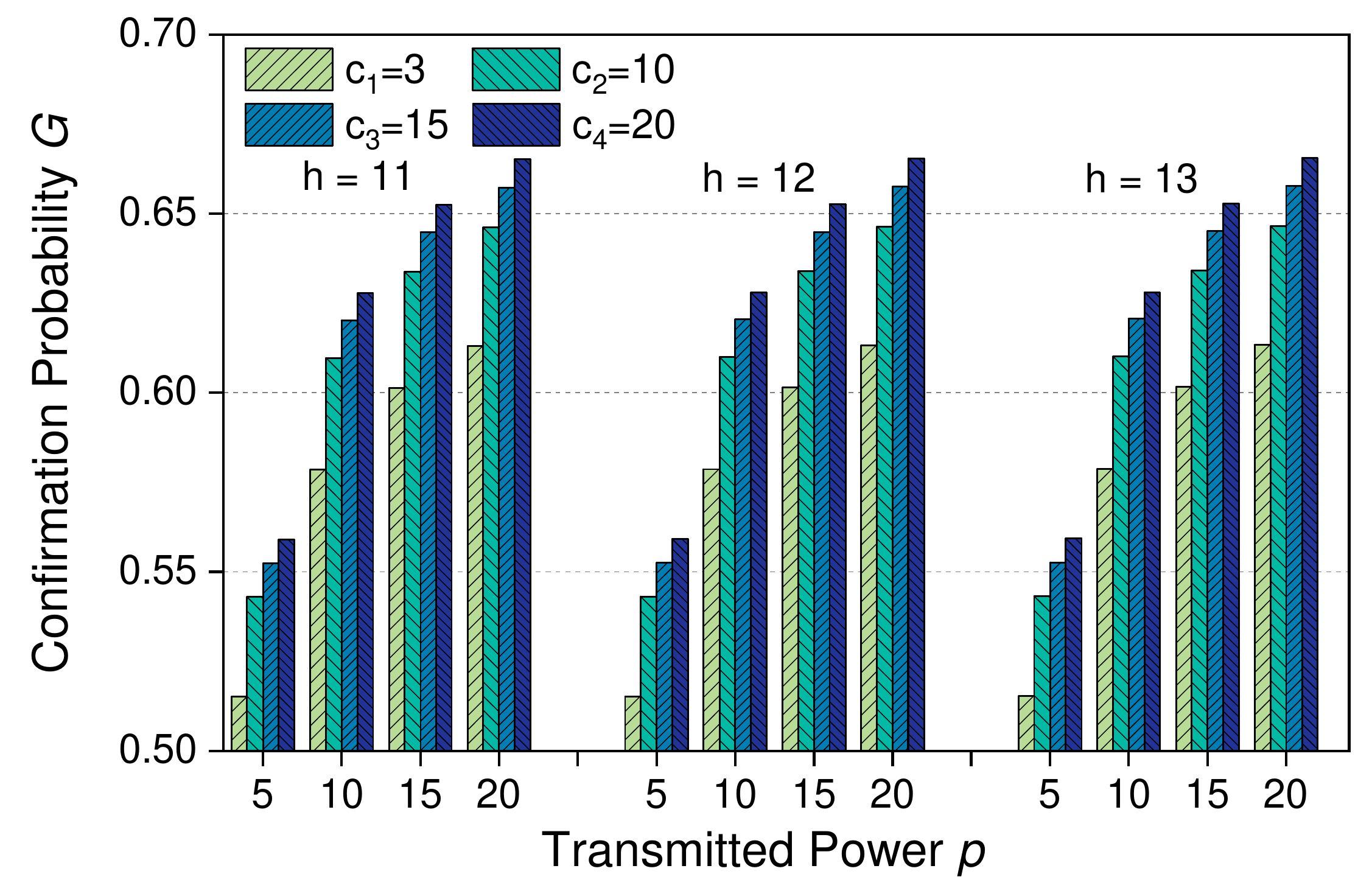}
  \caption{The impact of hash power $h$, connectivity $c$, and transmission power $p$ on
  the IoT device's confirmation probability $G$.}
\end{figure}

Then, we discuss the impact of the hash power $h$, connectivity
$c$, and transmission power $p$ on the IoT device's block confirmation 
probability. From Fig. 8, we observe the connectivity $c$ and transmission 
power $p$ have the dominant effect on the block confirmation probability 
$G$ compared with the hash power $h$, which is consistent with Section \mbox{III} B. 
Besides, the increase in transmission power $p$ has a more prominent effect 
than connectivity $c$ when both are large. When the connectivity $c$ is beyond 10, 
it has a slight effect on the block confirmation probability as the value increases. 
Similarly, the impact on confirmation probability is also finite as the transmission 
power $p$ increases. On the one hand, increasing the computing and communication capability 
improves the block confirmation probability. On the other hand, more powerful 
computing and communication capability bring greater energy consumption and limited improvement 
of confirmation probability. Therefore, it is significant for IoT devices to trade off computing and communication capability improvement.

\subsection{Feasibility of the Proposed Contract}

In Fig. 9, we present the rewards of IoT devices. The olive green
bar represents the fixed salary $s$, and the light blue bar represents
the bonus $B$. Both the $s$ and $B$ increase with the type
$\lambda$, which means the more capable devices contribute more effort
and gain more profits. It is consistent with the monotonicity constraint
and conforms to the law of economic life. 
From Fig. 9, we find that the blockchain developer prefers devices with 
more powerful capabilities and provides exponentially increasing rewards.

\begin{figure}[t]
  \includegraphics[scale=0.39]{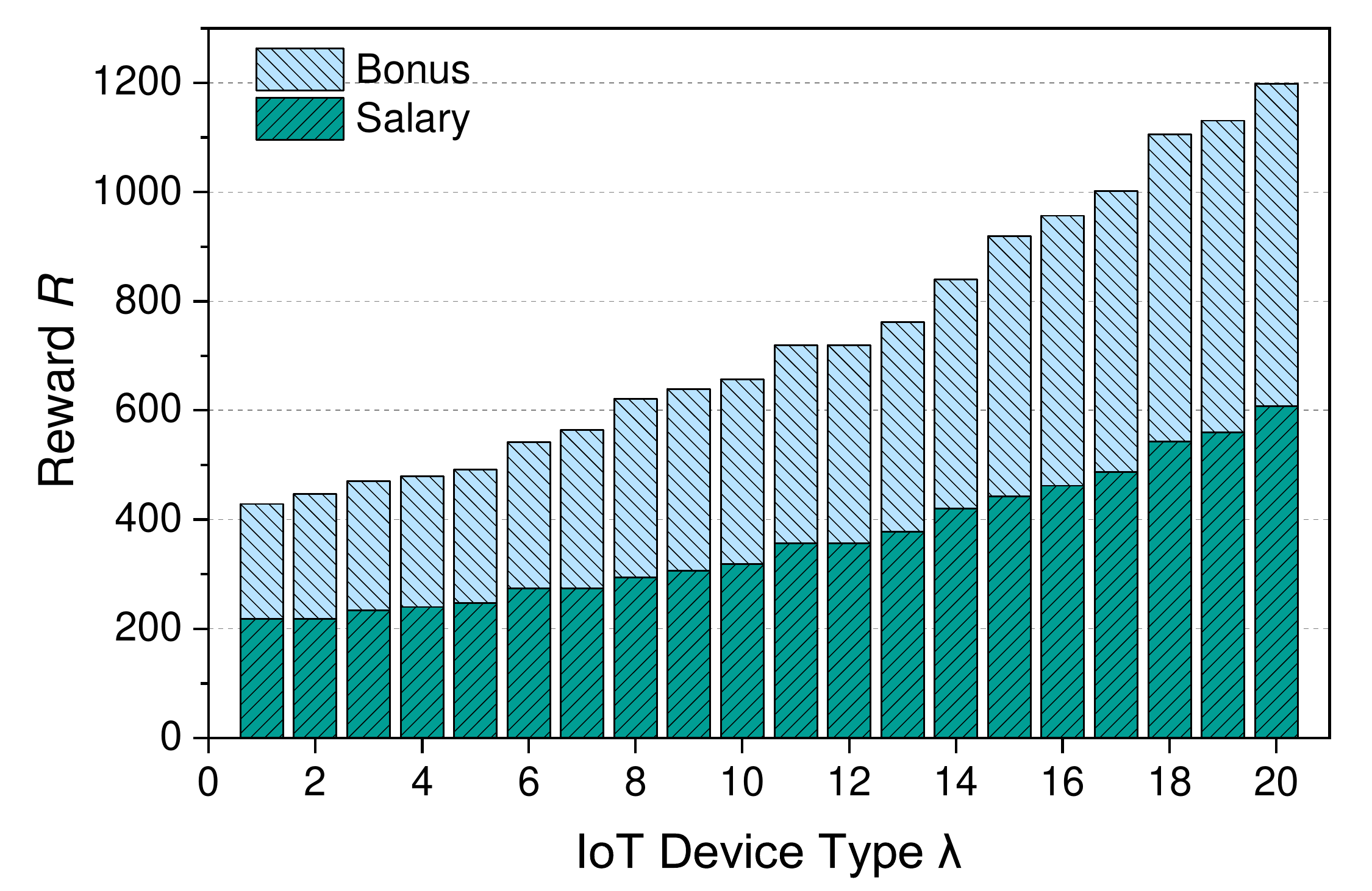}
  \caption{The reward $R$ for each IoT device's type.}
\end{figure}

\begin{figure}[t]
  \includegraphics[scale=0.39]{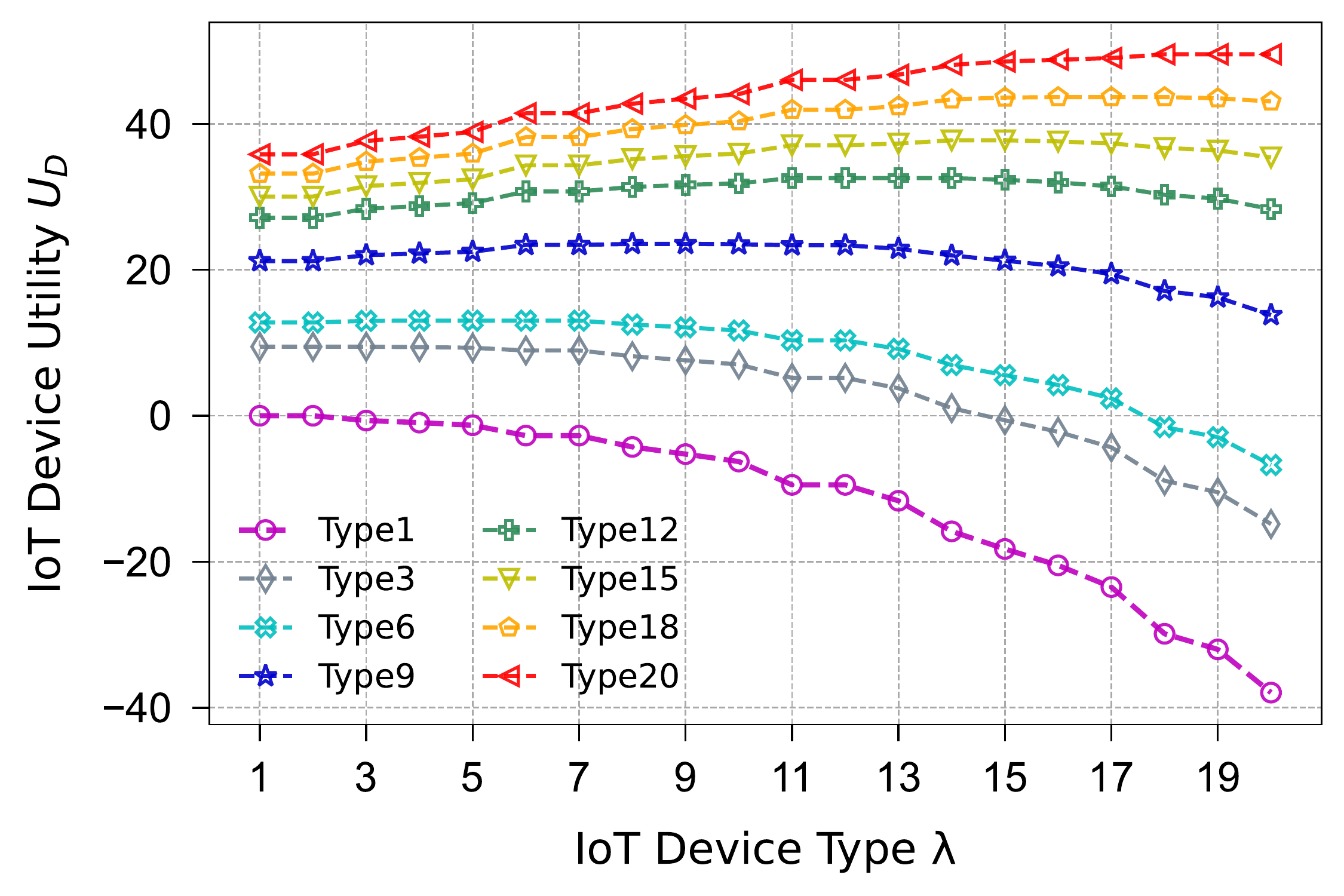}
  \caption{The feasibility of the proposed contract.}
\end{figure}

Fig. 10 illustrates the feasibility of our proposed contract. From
Fig. 10, we observe that the IoT device gains the non-negative utility,
which satisfies the IR constraints. The utility of type $\lambda_{1}$ device
is set to be zero and consistent with the equation \eqref{eq:one IR-1}.
Furthermore, the IoT device gains maximum utility while selecting
the contract of the corresponding type, which satisfies the IC constraints.

We compare the proposed contract
with the contract that does not consider the block confirmation probability, which
is the \emph{contract only with adverse selection} \cite{Xiong2020}. 
To demonstrate the advantage of contract theory, we add a comparison to the 
existing PoW-based incentive mechanism, where every device obtains the same rewards after building a 
block. We set the median utility of heterogeneous devices as a fixed incentive.
Additionally, we compare our contract with the perfect information contract, 
where the developer knows the actual capability of each IoT 
device. In the perfect information scenario, the developer elaborates the contract 
for every IoT device and ensures that they obtain zero utility. The wireless blockchain 
utility with perfect information contract is: 

\begin{subequations}
\begin{align}
\max_{(s_{i},b_{i})}U_{B}= & \sum_{i=1}^{L}ZQ(\lambda_{i})\Bigl(2\varepsilon\lambda_{i}b_{i}-s_{i}-2\lambda_{i}b_{i}^{2}\Bigr),\label{eq:one-dimensional contract-1}\\
s.t. & \ \ U_{D}^{i}(\omega_{i})=0.
\end{align}
\end{subequations}

Thus, the developer can maximize the wireless blockchain utility with a perfect information contract. 
In Fig. 11, we compare the proposed contract and the approach in \cite{Xiong2020} over the wireless 
blockchain utility. While the yield 
coefficient $\varepsilon$ varies from $100$ to $550$, the wireless blockchain 
utility of the proposed contract is about $35\%$ higher than the approach in \cite{Xiong2020} 
and increases utility by 4 times compared with the original PoW-based incentive mechanism.
We also can see that our contract is very close to the perfect information contract, 
which demonstrates the effectiveness of our contract. 

\begin{figure}[t]
  \includegraphics[scale=0.39]{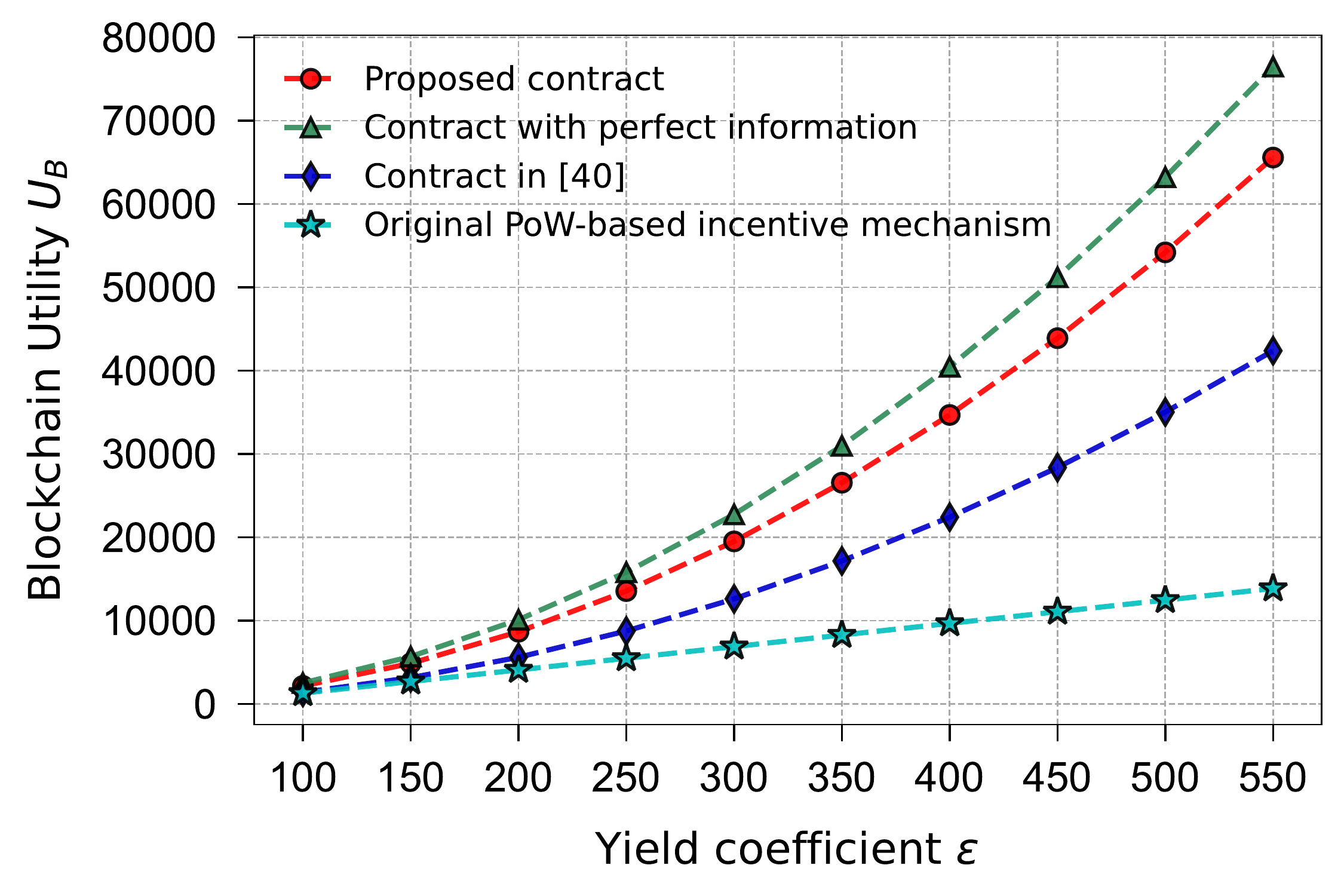}
   \caption{The blockchain utility under different incentive mechanisms.}
\end{figure}

\section{Conclusions and Future Work}
In this paper, we have investigated the problem of incentive mechanism
design in the  wireless blockchain IoT network to maximize the wireless 
blockchain utility. Taking into account the IoT device's hash power, connectivity, and 
transmission power, we analyze IoT devices' energy consumption and block confirmation 
probability while maintaining the wireless blockchain network. We propose a multi-dimensional
contract to address the \emph{adverse selection} and \emph{moral hazard} issues, 
and attract more IoT devices to join the wireless blockchain network. In particular, 
we explore the network connectivity's effect on the block confirmation probability from 
the complex wireless network perspective. Furthermore, we find that connectivity affects the 
confirmation probability dissimilarly over different network sizes and average link probability. 
From the simulation results, our proposed contract is effective and closer to the 
theoretical optimal utility than the contract with adverse selection. In addition, the communication factors 
have a more prominent effect on block confirmation than hash power in the wireless blockchain 
network, which is different from the traditional blockchain networks. 

In future work, we will explore the Integration of blockchain and artificial 
intelligence (AI) to construct a trusted environment for next-generation computing.
Blockchain has much potential to be combined with 
fog/edge/serverless and quantum computing scenarios \cite{gill2022ai}. In cloud/fog/edge computing, 
blockchain provides a platform for both server and user to allocate resources efficiently. 
From the perspective of serverless computing, blockchain can replace traditional servers to converge 
AI models in federated learning (FL). A completely decentralized, secure, efficient, and privacy-protecting
machine learning framework could be achieved by introducing the blockchain. Besides, with the 
development of quantum computing, quantum-resistant encryption algorithms need to be further investigated 
to ensure the security of blockchain systems. The ability to form Decentralized Autonomous 
Organizations (DAOs) is a concept fundamental to the Cloud-to-Things computing continuum, which is 
consistent with the decentralized nature of blockchain.


\ifCLASSOPTIONcaptionsoff
  \newpage
\fi



%


\bibliographystyle{IEEEtran}
\bibliography{citation}

%





\end{document}